%% file: ms.tex
\documentclass[lettersize,journal]{IEEEtran}
\pdfoutput=1
\usepackage{amsmath,amsfonts}
\usepackage{algorithmic}
\usepackage{algorithm}
\usepackage{array}
\usepackage[caption=false,font=normalsize,labelfont=sf,textfont=sf]{subfig}
\usepackage{textcomp}
\usepackage{stfloats}
\usepackage{url}
\usepackage{verbatim}
\usepackage{graphicx}
\usepackage{lineno}
\usepackage{cite}
\begin{document}

\title{Bayesian seismic tomography based on velocity-space Stein variational gradient descent for physics-informed neural network}

\author{Ryoichiro Agata, Kazuya Shiraishi and Gou Fujie
        % <-this % stops a space
%\thanks{This paper was produced by the IEEE Publication Technology Group. They are in Piscataway, NJ.}% <-this % stops a space
\thanks{\copyright 2023 IEEE.  Personal use of this material is permitted.  Permission from IEEE must be obtained for all other uses, in any current or future media, including reprinting/republishing this material for advertising or promotional purposes, creating new collective works, for resale or redistribution to servers or lists, or reuse of any copyrighted component of this work in other works. DOI: 10.1109/TGRS.2023.3295414}
\thanks{Ryoichiro Agata, Kazuya Shiraishi and Gou Fujie are with Japan Agency for Marine-Earth Science and Technology (JAMSTEC). }}

%\markboth{This manuscript was submitted to IEEE Transactions on Geoscience and Remote Sensing}%
%\markboth{This work has been submitted to the IEEE for possible publication. Copyright may be transferred without notice, \\after which this version may no longer be accessible.}
\markboth{This manuscript was accepted for IEEE Transactions on Geoscience and Remote Sensing}%
{Shell \MakeLowercase{\textit{et al.}}: A Sample Article Using IEEEtran.cls for IEEE Journals}


\maketitle
%\IEEEpeerreviewmaketitle

%\linenumbers

\begin{abstract}
\input{abst.tex}
\end{abstract}


\input{sct1.tex}

\input{sct2.tex}

\input{sct2.5.tex}

\input{sct3.tex}

\input{sct4.tex}

\input{sct5.tex}

\input{sct6.tex}

\section*{Acknowledgments}
\input{ack.tex}

\input{algo.tex}

\input{fig.tex}

\input{tab.tex}

\input{app1.tex}


\input{IEEE_style.bbl}

%\bibliographystyle{IEEEtran}
%\bibliography{C:/Users/sgbea/ownCloud/commonsource/latex/bibtex/bibtex_agata}

%\newpage

\section{Biography Section}
% your own custom macro containing the $\backslash${\tt{includegraphics}} command to make things
 

%\bf{If you include a photo:}\vspace{-33pt}
%\begin{IEEEbiography}[{\includegraphics[width=1in,height=1.25in,clip,keepaspectratio]{fig/fig1.png}}]{Ryoichiro Agata}
%$\backslash${\tt{begin\{IEEEbiography\}}} and then for the 1st argument use $\backslash${\tt{includegraphics}} to declare and link the author photo.
%Use the author name as the 3rd argument followed by the biography text.
%\end{IEEEbiography}


\begin{IEEEbiographynophoto}{Ryoichiro Agata}
Ryoichiro Agata is a Researcher with Japan Agency for Marine-Earth Science and Technology (JAMSTEC).
\end{IEEEbiographynophoto}

\begin{IEEEbiographynophoto}{Kazuya Shiraishi}
Kazuya Shiraishi is a Researcher with Japan Agency for Marine-Earth Science and Technology (JAMSTEC).
\end{IEEEbiographynophoto}

\begin{IEEEbiographynophoto}{Gou Fujie}
Gou Fujie is Director of Subduction Dynamics Research Center, Research Institute for Marine Geodynamics at Japan Agency for Marine-Earth Science and Technology (JAMSTEC).
\end{IEEEbiographynophoto}

\end{document}

%% file: abst.tex
In this study, we propose a Bayesian seismic tomography inference method using physics-informed neural networks (PINN). 
PINN represents a recent advance in deep learning, offering the possibility to enhance physics-based simulations and inverse analyses. 
PINN-based deterministic seismic tomography uses two separate neural networks (NNs) to predict seismic velocity and travel time. 
Naive Bayesian NN (BNN) approaches are unable to handle the high-dimensional spaces spanned by the weight parameters of these two NNs. 
Hence, we reformulate the problem to perform the Bayesian estimation exclusively on the NN predicting seismic velocity, while the NN predicting travel time is used only for deterministic travel time calculations, with the help of the adjoint method. 
Furthermore, we perform BNN by introducing a function-space Stein variational gradient descent (SVGD), which performs particle-based variational inference in the space of the function predicted by the NN (i.e., seismic velocity), instead of in the traditional weight space. 
The result is a velocity-space SVGD for the PINN-based seismic tomography model (vSVGD-PINN-ST) that decreases the complexity of the problem thus enabling a more accurate and physically consistent Bayesian estimation, as confirmed by synthetic tests in one- and two-dimensional tomographic problem settings. 
The method allows PINN to be applied to Bayesian seismic tomography practically for the first time. 
Not only that, it can be a powerful tool not only for geophysical but also for general PINN-based Bayesian estimation problems associated with compatible NNs formulations and similar, or reduced, complexity.

%% file: sct1.tex
\section{Introduction}
\label{sct:intro}

Seismic tomography is a technique used to determine the interior seismic structure of the Earth, using seismic waves excited by natural earthquakes and artificial sources. This technique is essential for studying the evolution of the Earth, plate tectonics, and mechanisms of earthquake generation. 
 Seismic tomography, similarly to many other physics-based estimation problems in geoscience, represents an under-determined inverse problem because the number of sources and receivers is limited. 
A priori constraints on the parameters of interest (e.g., seismic velocity) can be used to regularize the problem and provide the uncertainty quantification (UQ) of the estimates, ensuring reliable analytical results. 
To meet such requirement, recently, a growing body of research uses Bayesian estimation to perform UQ estimation, considering unknown parameters as stochastic variables. 
In the field of seismic tomography, there are many examples of Bayesian estimations in two-dimensional (2D) surface wave \cite{Bodin2009,Bodin2012,Galetti2015,Zhang2020Seismic}, 2D seismic refraction \cite{Ryberg2018}, and three-dimensional (3D) tomography \cite{Piana2015,Hawkins2015,Burdick2017}. 
These methods employ a Bayesian estimation of the posterior probability distribution of the seismic velocity structure, combining travel time data, prior information on the target velocity structure, and forward numerical calculations of travel time based on grid or mesh discretization of the target domain.

Recent developments in deep learning techniques provide entirely new options for solving partial differential equations (PDEs) and inversion problems. 
Physics-informed neural networks (PINN) \cite{Raissi2019}, in particular, have attracted considerable attention owing to their high applicability and flexibility. 
Raissi et al. \cite{Raissi2019} presented a PINN application to Navier--Stokes equations. It now has been applied to a variety of fields, including solid earth science, in forward simulations (e.g., seismic travel time calculation \cite{Smith2020,Waheed2021PINNeik}, seismic wave field simulation \cite{Song2022}, and crustal deformation modeling \cite{Okazaki2022}), and inverse problems (e.g., seismic tomography \cite{Waheed2021PINNtomo,Chen2022} and full-waveform inversion \cite{Rasht-Behesht2022}).
PINN uses neural networks (NN) to construct functions predicting physical quantities, such as travel time, taking spatial coordinates and time (for dynamic problems) as inputs.
The NN is trained based on a loss function consisting of residuals of the governing PDE in some given evaluation (or collocation) points. 
The advantages brought by the PINN travel time calculation \cite{Smith2020,Waheed2021PINNeik} and seismic tomography in general \cite{Waheed2021PINNtomo,Chen2022}, such as mesh-free frameworks not requiring initial model setup, demonstrate a considerable potential for further important developments.
The one considered in this study aims at extending PINN-based seismic tomography to include UQ based on Bayes' theorem.

Bayesian inversion based on PINN is an application of Bayesian neural network (BNN) where the posterior probability density function (PDF) of the weight parameters in NN is estimated. 
Stochastic properties of target physical quantities, such as seismic wave velocity, are obtained from the posterior predictive PDF, calculated based on the posterior PDF. 
Bayesian sampling methods and variational inference (VI) are the methods most commonly used to estimate the posterior PDF of the weight parameters in BNN \cite{Gawlikowski2021}. 
Bayesian sampling is the most accurate method for BNN, enabling sampling from the true posterior PDF.
For instance, the Hamiltonian/hybrid Monte Carlo (HMC) method \cite{Duane1987} has been used in previous applications of Bayesian PINN \cite{Yang2021,Linka2022,Psaros2023}. 
However, being a sequential sampling method, HMC computation time and scalability are not suitable for large problems.
In VI, the posterior PDF estimation problem is replaced with an optimization one, minimizing the Kullback--Leibler (KL) divergence between the target PDF and its parametric approximation. 
VI are usually based on the mean-field approximation (MFVI), which approximates the target PDF combining multiple independent simple PDFs. 
However, MFVI often displays poor performance in Bayesian PINN, giving an over-simplified PDF approximation \cite{Yang2021,Psaros2023}. 
A more recent non-parametric class of VI, called particle-based VI (ParVI) and best known as Stein variational gradient descent (SVGD) \cite{Liu2016}, has attracted attention for the characteristic high approximation accuracy and computational parallelism. 
ParVI methods iteratively update a set of particles and use the corresponding empirical probability measure to approximate the target posterior PDF accurately.
These methods have already been applied both to BNN and Bayesian PINN problems, using relatively small NN structures (e.g., \cite{Sun2020}). 
Even so, BNNs have a challenging aspect, that is, the multi-modality of the posterior PDF in the high-dimensional weight space resulting from over-parametrization. 
In fact, optimized networks are able to express the same function values using multiple parameter combinations. 
Exploring PDFs solutions using ParVI and SVGD in such an over-parametrized space is a difficult task, resulting in degraded approximation performance \cite{Wang2019}.

However, in physics problems, such as those of seismic tomography, the posterior PDFs of interest are defined in the space of the functions predicted by NNs (e.g., seismic velocities); therefore, they are expected to have a considerably simpler shape than those defined in the weight space. 
Performing ParVI in function space, instead of weight space with multi-modality, in BNN avoids posterior PDF degraded approximation issues \cite{Wang2019}.
In addition, considering Bayesian estimations in the function space also enables the efficient incorporation of physically meaningful components of prior information, which would not be possible in the weight space \cite{Sun2019}.

Another important aspect to consider is the confinement of a NN targeted in Bayesian estimation. 
In fact, previous PINN-based seismic tomography methods \cite{Waheed2021PINNtomo,Chen2022} included two NNs: one predicting travel time as a solution of the governing PDE and the other seismic wave velocity. 
A similar NN formulation was employed in other inverse problems (e.g. \cite{Rasht-Behesht2022}). 
Targeting the weight parameters of two NNs in Bayesian estimation is not efficient, considering we are exclusively interested in the seismic velocity UQ. 
Moreover, the only requirement for the travel time NN is to satisfy the governing PDE for the given velocity structure. 
Therefore, if we could confine the target parameters to the weights of the NN predicting seismic velocity (or other relevant functions for UQ), the exploration of the posterior PDF would be drastically simplified.

This study proposes a novel method, based on Bayesian inference and PINN, to address UQ and inverse problems applied to seismic tomography, overcoming existing limitations. 
The main contributions of this study can be summarized as follows. 
We reformulated the problem to perform the Bayesian estimation only on the velocity NN, while using the travel time NN for deterministic travel time calculations, with the help of a numerical technique called the adjoint method \cite{Lewis1985}. 
We performed Bayesian estimation for the velocity network by introducing SVGD directly in the space of functions predicted by the NN (i.e., seismic velocity), instead of in the whole weight space, leveraging a recent development in BNN techniques \cite{Wang2019}. 
The proposed approach is called ``velocity-space SVGD for PINN-based seismic tomography (vSVGD-PINN-ST)'' and represents the first practical application of Bayesian estimation in PINN-based seismic tomography. 
Furthermore, vSVGD-PINN-ST can be generalized to the UQ of PINN-based inverse problems sharing similar NN formulation and problem size.

This paper is organized as follows: In Section \ref{sct:baseline}, we present a PINN formulation of seismic tomography from previous studies and explain a naive formulation of Bayesian estimation  as a baseline method. In Section \ref{sct:vSVGD-PINN}, we present our vSVGD-PINN-ST, introducing the improvements presented earlier. 
In Section \ref{sct:linear} and \ref{sct:2D}, we validate the applicability of the proposed method to realistic problems through synthetic seismic tomography tests in one and two dimensions. In Section \ref{sct:discussion} and \ref{sct:conclusion} we discuss results and give concluding remarks, respectively.

%% file: sct2.tex
\newcommand{\argmin}{\mathop{\rm arg~min}\limits}

\section{Baseline method}
\label{sct:baseline}

\subsection{PINN-based deterministic seismic tomography}
\label{sct:PINNtomo}

First, we first introduce the PINN formulation of the deterministic seismic tomographic problem as proposed by \cite{Waheed2021PINNtomo}. 
In this work, we follow a slightly modified approach. 
We start from the eikonal equation relating the spatial derivative of the travel time field to the velocity structure as follows:
\begin{eqnarray}
|\nabla T(\mathbf{x},\mathbf{x}_{s})|^{2} &=& \displaystyle \frac{1}{v^{2}(\mathbf{x})}, \quad \forall\,\mathbf{x} \in \Omega \label{eqn:eikonal}\\
T(\mathbf{x}_{s}, \mathbf{x}_{s}) &=& 0,\label{eqn:PC}
\end{eqnarray}
where $\Omega$ is a $\mathbb{R}^d$ domain, with $d$ as the space dimension, $T(\mathbf{x},\mathbf{x}_{s})$ is the travel time at the point $\mathbf{x}$ from the source $\mathbf{x}_s$, $v(\mathbf{x})$ is the velocity defined on $\Omega$, and $\nabla$ denotes the gradient operator. 
The second equation defines the point source condition. 
To avoid singularities in this condition, previous studies modeling the travel time using PINN introduced the following factored form \cite{Smith2020,Waheed2021PINNeik}: 
\begin{eqnarray}
T(\mathbf{x},\mathbf{x}_{s})=T_{0}(\mathbf{x},\mathbf{x}_{s}) \tau(\mathbf{x},\mathbf{x}_{s})
\end{eqnarray}
where $T_{0}(\mathbf{x})$ is defined as
\begin{eqnarray}
T_{0}(\mathbf{x},\mathbf{x}_{s})=\left|\mathbf{x}-\mathbf{x}_{s}\right|.
\end{eqnarray}
This factorization automatically satisfies the point source condition.
%becomes: 
%\begin{eqnarray}
%\tau(\mathbf{x}_{s},\mathbf{x}_{s})=\frac{1}{v(\mathbf{x}_{s})} %\label{eqn:PC_factored}. 
%\end{eqnarray}
Considering $v$ positive, we can introduce the residuals of the eikonal equation $r_{\rm EE}$ 
%and the point source condition $r_{\rm PC}$ 
in terms of velocity:
\begin{eqnarray}
r_{\rm EE}&=&v(\mathbf{x})-\frac{1}{|\nabla T(\mathbf{x},\mathbf{x}_{s})|},\label{eqn:residual}
%r_{\rm PC}&=&v(\mathbf{x}_{s})-\frac{1}{\tau(\mathbf{x}_{s},\mathbf{x}_{s})}\label{eqn:residual_PC}. 
\end{eqnarray}
The PINN-based seismic tomography method solves the eikonal equation and estimates the velocity structure simultaneously, by training NNs to predict the travel-time and velocity function. 
Waheed et al. \cite{Waheed2021PINNtomo} proposed to use two different NNs to construct the functions $f_{T}$ and $f_{v}$, characterized by weight parameters $\boldsymbol{\theta}_{T}$ and $\boldsymbol{\theta}_{v}$. 
Optimized NNs are expected to return accurate approximations of the travel time and velocity. Such formulation, which introduces two NNs, one for the solution of the governing PDE and the other for PDE parameters optimization, is generally used in PINN-based inverse analysis \cite{Rasht-Behesht2022}.
In this study, we define NNs functions as 
\begin{eqnarray}
T(\mathbf{x},\mathbf{x}_{s}) 
&\simeq& f_{T}(\mathbf{x},\mathbf{x}_{s},\boldsymbol{\theta}_{T})\nonumber\\
&=& T_{0}(\mathbf{x},\mathbf{x}_{s})/f_{\tau^{-1}}(\mathbf{x},\mathbf{x}_{s},\boldsymbol{\theta}_{T}),\label{eqn:T}\\
v(\mathbf{x})
&\simeq& f_{v}(\mathbf{x},\boldsymbol{\theta}_{v})\nonumber\\
&=& v_0(\mathbf{x}) + f_{v_{\rm ptb}}(\mathbf{x},\boldsymbol{\theta}_{v}),
\end{eqnarray}
where $f_{\tau^{-1}}$ is an NN-based function approximating $1/\tau(\mathbf{x},\mathbf{x}_{s})$, 
$v_0(\mathbf{x})$ is the reference velocity set by the user, and $f_{v_{\rm ptb}}(\mathbf{x},\boldsymbol{\theta}_{v})$ is the neural network function approximating the velocity perturbation component. 
We approximated $1/\tau$ and $v_{\rm ptb}$ using NNs, instead of directly computing $\tau$ and $v$ as done in previous studies \cite{Smith2020,Waheed2021PINNeik}, to improve convergence performances. 
We used fully connected feed-forward networks to implement both $f_{\tau^{-1}}$ and $f_{v_{\rm ptb}}$. 
The $f_{v_{\rm ptb}}$ network is characterized by the parameters $v_{\rm ptb}^{\rm max}$ and $v_{\rm ptb}^{\rm min}$, representing the maximum and minimum velocity, set a priori. 
Additional operations were applied to normalize the input and output of NNs, improving the convergence performance and setting upper and lower limits of the final output values.
See Appendix \ref{app:normalization} for details.
Further, the reciprocity condition (i.e., $T(\mathbf{x},\mathbf{x}_{s})=T(\mathbf{x}_{s},\mathbf{x})$) was imposed following \cite{Grubas2023} by using $\frac{1}{2}\left(f_{\tau^{-1}}(\mathbf{x},\mathbf{x}_{s},\boldsymbol{\theta}_{T})+f_{\tau^{-1}}(\mathbf{x}_{s},\mathbf{x},\boldsymbol{\theta}_{T})\right)$ instead of  $f_{\tau^{-1}}(\mathbf{x},\mathbf{x}_{s},\boldsymbol{\theta}_{T})$ in Equation  \ref{eqn:T} to improve the convergence of solution of the eikonal equation.   
For a deterministic tomographic problem, the two NNs are trained simultaneously using the same loss function:
\begin{eqnarray}
L 
&=& \alpha_1 \sum_{i=1}^{N_T}\left(T_{\rm obs}^{(i)}-f_{T}(\mathbf{x}_{r}^{(i)},\mathbf{x}_{s}^{(i)}; \boldsymbol{\theta}_{T})\right)^{2}\nonumber\\
&+& \alpha_2 \sum_{i=1}^{N_c}\left( f_{v}(\mathbf{x}_{c}^{(i)}; \boldsymbol{\theta}_{v})-\frac{1}{|\nabla f_{T}(\mathbf{x}_{c}^{(i)},\mathbf{x}_{s}^{(i)}; \boldsymbol{\theta}_{T})|} \right)^{2},
%&+& \alpha_3 \sum_{i=1}^{N_s}\left( f_{v}(\mathbf{x}_{s}^{(i)}; \boldsymbol{\theta}_{v}) - f_{\tau^{-1}}(\mathbf{x}_{s}^{(i)},\mathbf{x}_{s}^{(i)};\boldsymbol{\theta}_{T}) \right)^{2},
\end{eqnarray}
where $N_T$, $N_c$ and $N_s$ are the number of travel time data, collocation points, and source points, respectively;
$T_{\rm obs}$ represents the travel time data; 
$\mathbf{x}_{r}$ and $\mathbf{x}_{c}$ are the coordinates of receiver and collocation points, respectively; $\alpha_1$, $\alpha_2$, and $\alpha_3$ are loss weights, typically assigned to 
$1/{N_{T}}$, $1/{N_{c}}$ and $1/{N_{s}}$, respectively, although manual or automated tuning can be used for a stable training \cite{Wang2022}. 
The collocation points, which are usually selected randomly within the target domain $\Omega$, are set as evaluation points of the PDE residuals \cite{Raissi2019}. 
The first term on the right-hand side is the loss due to observational constraints, consisting of the sum of the squared difference of true and predicted travel time values for each source--receiver pair.
The second term is the loss due to physics-informed constraints, consisting of the sum of the squared residuals defined in Equation \ref{eqn:residual} for each pair of source and collocation point. 
%The third term is the one modeling the point source constraints, consisting of the sum of the squared residual defined in Equation \ref{eqn:residual_PC} for each source point. 
 
Following the above definitions, we defined the training datasets for the observation and PDE loss functions as 
${\bf X}_T = \{ ({\bf x}_r^{(1)}, {\bf x}_s^{(1)}, T_{\rm obs}^{(1)}),({\bf x}_r^{(2)}, {\bf x}_s^{(2)}, T_{\rm obs}^{(2)}),\dots,({\bf x}_r^{(N_T)}, {\bf x}_s^{(N_T)}, T_{\rm obs}^{(N_T)}) \}$ and 
${\bf X}_c = \{ ({\bf x}_c^{(1)}, {\bf x}_s^{(1)}),({\bf x}_c^{(2)}, {\bf x}_s^{(2)}),\dots,({\bf x}_c^{(N_c)}, {\bf x}_s^{(N_c)}) \}$, respectively. 
Tomographic estimation of $v(\mathbf{x})$ is given by $f_{v}(\mathbf{x},\boldsymbol{\theta}_{v}^{*})$, where $\boldsymbol{\theta}_{T}^{*}, \boldsymbol{\theta}_{v}^{*}= \argmin_{\boldsymbol{\theta}_{T}, \boldsymbol{\theta}_{v}} L(\boldsymbol{\theta}_{T}, \boldsymbol{\theta}_{v})$ represent the NNs optimized values. 
Fig. \ref{fig:NN_shematics} schematically illustrates settings and constraints used to train the NNs involved in both the tomographic deterministic and Bayesian formulation (described in the following section).

\subsection{Naive Bayesian formulation of PINN-based seismic tomography}

In this study, we considered the weight parameters as stochastic variables, formulating the estimation problem using the Bayes' theorem. 
In a naive formulation, we consider a Bayesian neural network (BNN) for both $\boldsymbol{\theta}_{T}$ and $\boldsymbol{\theta}_{v}$ given by:
\begin{eqnarray}
P(\boldsymbol{\theta}_{T},\boldsymbol{\theta}_{v}|\mathbf{d}) &=& \frac{P(\mathbf{d} | \boldsymbol{\theta}_{T},\boldsymbol{\theta}_{v}) P(\boldsymbol{\theta}_{T},\boldsymbol{\theta}_{v})}{P(\mathbf{d})}\nonumber\\ 
&\propto& P(\mathbf{d} | \boldsymbol{\theta}_{T},\boldsymbol{\theta}_{v}) P(\boldsymbol{\theta}_{T},\boldsymbol{\theta}_{v}),\label{eqn:bayes}
\end{eqnarray}
where $\mathbf{d}$ is the data vector, 
including not only travel time data but also the constraints from the eikonal equation:
The likelihood function includes the two following components: 
\begin{eqnarray}
P(\mathbf{d} | \boldsymbol{\theta}_{T},\boldsymbol{\theta}_{v})=P({\bf T}_{\rm obs} | \boldsymbol{\theta}_{T},\boldsymbol{\theta}_{v}) P({\bf r}| \boldsymbol{\theta}_{T},\boldsymbol{\theta}_{v}). 
\label{eqn:lh}
\end{eqnarray}
The first term on the right-hand side represents the stochastic observation error of the travel time, for which the following Gaussian distribution is assumed: 
\begin{align}
&P({\bf T}_{\rm obs} | \boldsymbol{\theta}_{T},\boldsymbol{\theta}_{v})\nonumber\\
&=\frac{1}{Z}\exp\left( -\frac{1}{2}({\bf T}_{\rm obs}-{\bf f}_{T}(\boldsymbol{\theta}_{T}))^{\top} {\bf E}^{-1}_{\rm obs} ({\bf T}_{\rm obs}-{\bf f}_{T}(\boldsymbol{\theta}_{T})) \right),
\label{eqn:lh_TT}
\end{align}
where
\begin{eqnarray}
{\bf T}_{\rm obs} &=& \left[T_{\rm obs}^{(1)}, T_{\rm obs}^{(2)},\dots,T_{\rm obs}^{(N_T)} \right],\\
{\bf f}_{T}(\boldsymbol{\theta}_{T}) &=& \left[ f_T^{(1)},f_T^{(2)},\dots,f_T^{(N_T)} \right],\\
f_T^{(i)} &=& f_{T}(\mathbf{x}_{r}^{(i)},\mathbf{x}_{s}^{(i)}; \boldsymbol{\theta}_{T}),
\end{eqnarray}
and ${\bf E}_{\rm obs}$ and $Z$ are the covariance matrix for the observation error and a normalizing coefficient, respectively. 
We see that $P({\bf T}_{\rm obs} | \boldsymbol{\theta}_{T},\boldsymbol{\theta}_{v})$ is an explicit function of $\boldsymbol{\theta}_{T}$, implicitly dependent on $\boldsymbol{\theta}_{v}$. 
To materialize the second term of Equation \ref{eqn:lh}, we assume Gaussian properties for the residuals as well: 
\begin{align}
&P({\bf r} | \boldsymbol{\theta}_{T},\boldsymbol{\theta}_{v})\nonumber\\
&=\frac{1}{Z}\exp\left( -\frac{1}{2}{\bf r}^{\top}(\boldsymbol{\theta}_{T}, \boldsymbol{\theta}_{v}) {\bf E}^{-1}_r {\bf r}(\boldsymbol{\theta}_{T}, \boldsymbol{\theta}_{v}) \right),
\label{eqn:lh_pde}
\end{align}
where 
\begin{align}
&{\bf r}(\boldsymbol{\theta}_{T}, \boldsymbol{\theta}_{v}) = \left[ r_{\rm EE}^{(1)}, r_{\rm EE}^{(2)},\dots, r_{\rm EE}^{(N_c)} \right],\\ 
&r_{\rm EE}^{(i)} = f_{v}(\mathbf{x}_{c}^{(i)}; \boldsymbol{\theta}_{v})-\frac{1}{|\nabla f_{T}(\mathbf{x}_{c}^{(i)},\mathbf{x}_{s}^{(i)}; \boldsymbol{\theta}_{T})|},
\end{align}
%, r_{\rm PC}^{(1)}, r_{\rm PC}^{(2)},\dots, r_{\rm PC}^{(N_s)} \right],\\ 
%&r_{\rm PC}^{(i)} = f_{v}(\mathbf{x}_{s}^{(i)}; \boldsymbol{\theta}_{v}) - f_{\tau^{-1}}(\mathbf{x}_{s}^{(i)},\mathbf{x}_{s}^{(i)};\boldsymbol{\theta}_{T}),
and ${\bf E}_{\rm r}$ is the covariance matrix of PDE residuals, which may only have diagonal components given depending on the confidence level of the forward model. 
We use the notations ${\bf r}(\boldsymbol{\theta}_{T}, \boldsymbol{\theta}_{v})$ and ${\bf r}$ intermittently.
The simplest choices for the prior PDF $P(\boldsymbol{\theta}_{T},\boldsymbol{\theta}_{v})$ are, for instance, an independent and identically distributed (i.i.d.) zero-mean Gaussian distribution \cite{Yang2021} or a student's $t$-distribution \cite{Sun2020}.
Once the specific forms of the likelihood function and prior PDF are determined, the approximate posterior PDF can be obtained by several Bayesian estimation methods, introduced in the next section. 
The stochastic property of the seismic velocity is then obtained as the predictive PDF $P({\bf v}|\mathbf{d})$ based on the marginal posterior PDF of $\boldsymbol{\theta}_{v}$ as follows:
\begin{eqnarray}
P({\bf v}|\mathbf{d}) &=& \int P({\bf v}|\boldsymbol{\theta}_{v})P(\boldsymbol{\theta}_{v}|\mathbf{d}) d\boldsymbol{\theta}_{v}.
\end{eqnarray}

\subsection{Stein variation gradient descent for naive Bayesian PINN-based seismic tomography}

The naive Bayesian formulation involves the Bayesian estimation of the weight parameters posterior PDF $P(\boldsymbol{\theta}_{T},\boldsymbol{\theta}_{v}|\mathbf{d})$. 
ParVI methods, such as SVGD \cite{Liu2016}, are paradigms of Bayesian estimation, recently gaining more popularity in various fields, including geophysics research (e.g., \cite{Zhang2020Seismic,Zhang2020Variational,Smith2022}). 
SVGD is known for a more efficient approximation ability in the calculation of the posterior distribution compared to HMC, which has degraded performance with high dimensional problems and large datasets.
As in ordinary MFVI, in SVGD the Bayesian estimation is replaced with a minimization problem of the KL divergence, defined as follows: 
\begin{eqnarray}
KL\left(Q(\boldsymbol{\theta}) \| P(\boldsymbol{\theta}|\mathbf{d})\right)=\int Q(\boldsymbol{\theta}) \log \frac{Q(\boldsymbol{\theta})}{P(\boldsymbol{\theta}|\mathbf{d})} d \boldsymbol{\theta},
\end{eqnarray}
where $P(\boldsymbol{\theta}|\mathbf{d})$ and $Q(\boldsymbol{\theta})$ are the target posterior and approximate distributions, respectively. 
SVGD employs a set of particles $\lbrace \boldsymbol{\theta} \rbrace ^n_{i=1}$ to approximate the target posterior PDF by minimizing the KL divergence. These particles iteratively move towards the posterior distribution following the gradient of the KL divergence $\boldsymbol{\varphi}$, which is obtained from the kernelized Stein discrepancy defined in a reproducing kernel Hilbert space (RKHS). 
The update equations of SVGD are given as follows:
\begin{eqnarray}
\boldsymbol{\theta}_{i}^{l+1}=\boldsymbol{\theta}_{i}^{l}+\epsilon_{l} \boldsymbol{\phi}\left(\boldsymbol{\theta}_{i}^{l}\right),
\label{eqn:SVGD_update}
\end{eqnarray}
where
\begin{eqnarray}
\boldsymbol{\phi}(\boldsymbol{\theta})=\frac{1}{n} \sum_{j=1}^{n} \lbrace \underset{\rm Driving\,force}{\underline{k(\boldsymbol{\theta}_{j}^{l},\boldsymbol{\theta}) \nabla_{\boldsymbol{\theta}_{j}^{l}}\log P(\boldsymbol{\theta}_{j}^{l}|{\bf d})}}+\underset{\rm Repulsive\,force}{\underline{\nabla_{\boldsymbol{\theta}_{j}^{l}} k(\boldsymbol{\theta}_{j}^{l},\boldsymbol{\theta})}} \rbrace,
\label{eqn:SVGD_vector}
\end{eqnarray}
$\epsilon_l$ is the step size at each iteration $l$, which can be determined using various adaptive optimizers (such as Adam \cite{Kingma2015}), and $k(\mathbf{x}, \cdot)$ represents a positive definite kernel. 
In particular, we adopt a radial basis function (RBF) kernel with bandwidth determined by the median heuristic as in previous studies. 
The ``driving force'' term is a smoothed gradient of the log posterior density that moves the particles toward the high-density regions of the posterior distribution. The ``repulsive force'' term promotes diversity and prevents particles from concentrating on the mode of the target PDF. 
This combination of two forces results in an efficient non-parametric approximation of the posterior PDF, using a finite number of particles. 
In an SVGD optimization, a mini-batch stochastic gradient descent can be used \cite{Liu2016} for efficient optimization with large training datasets. 
In BNN applications, NNs corresponding to the particles $\lbrace \boldsymbol{\theta} \rbrace ^n_{i=1}$ are 
trained simultaneously by SVGD. 
SVGD can be applied to the naive Bayesian formulation of the target inverse problem introduced in the previous section by taking $\boldsymbol{\theta}=(\boldsymbol{\theta}_{T}\,\boldsymbol{\theta}_{v})^{\top}$.
We call this naive approach ``SVGD for PINN-based seismic tomography'' (SVGD-PINN-ST). 
SVGD-PINN-ST is classified as a ParVI-based approach of Bayesian PINN (e.g. \cite{Sun2020}). 
Algorithm \ref{alg:naive} summarizes the above baseline SVGD-PINN-ST algorithm with a mini-batch stochastic descent. 

In the application examples shown later, we only compare the proposed method with this ParVI-based approach and do not compare with other Bayesian PINN approaches mentioned in Section \ref{sct:intro} for the following reasons: 
The HMC-based approach has never been applied to realistic problems such as involving two neural networks with large sizes at a certain level aimed at 2D problems as in this study, but only been applied to ideal or small problems (e.g., \cite{Yang2021,Linka2022,Psaros2023}). It has been reported that the MFVI-based approach is often not sufficiently accurate (e.g., \cite{Yang2021,Psaros2023}). 

\section{Formulation of velocity-space SVGD for PINN-based seismic tomography}
\label{sct:vSVGD-PINN}

\subsection{Formulation of Bayesian estimation exclusive for velocity NN}
\label{sct:adjoint}

Although SVGD is applicable to BNN or Bayesian PINN theoretically, the parameter space $\boldsymbol{\theta}=(\boldsymbol{\theta}_{T}\,\boldsymbol{\theta}_{v})^{\top}$ of a practical problem setting is usually high-dimensional (above $10^{4}$ parameters), making the problem intractable for both HMC and SVGD. 
Consequently, reducing the effective parameter space dimension in the target Bayesian estimation is essential. 
For this reason, we seek to reduce the target space from $(\boldsymbol{\theta}_{T}\,\boldsymbol{\theta}_{v})^{\top}$ to $\boldsymbol{\theta}_{v}$, considering that 
we are exclusively interested in the uncertainty of $v$.
The only requirement for $\boldsymbol{\theta}_{T}$ is to minimize the residuals of the eikonal equation. 
We reformulate the Bayesian estimation defined in Equation \ref{eqn:bayes} by considering exclusively the estimation of the posterior PDF of $\boldsymbol{\theta}_{v}$ as follows:
\begin{eqnarray}
P(\boldsymbol{\theta}_{v}|\mathbf{d}) &\propto& 
P(\mathbf{T}_{\rm obs} | \boldsymbol{\theta}_{v}) P(\boldsymbol{\theta}_{v}),\label{eqn:bayes_short}
\end{eqnarray}
with the condition
\begin{eqnarray}
{\bf r}(\boldsymbol{\theta}_{T}, \boldsymbol{\theta}_{v})={\bf 0}.
\end{eqnarray}
To estimate the posterior probability in Equation \ref{eqn:bayes_short}, we consider an SVGD update for $\boldsymbol{\theta}_{v}$ only:
\begin{eqnarray}
\boldsymbol{\theta}_{v\,i}^{l+1}=\boldsymbol{\theta}_{v\,i}^{l}+\epsilon_{l} \boldsymbol{\phi}\left(\boldsymbol{\theta}_{v\,i}^{l}\right),\label{eqn:SVGD_update_improved1}
\end{eqnarray}
where
\begin{align}
&\boldsymbol{\phi}(\boldsymbol{\theta}_{v})\nonumber\\
&=\frac{1}{n} \sum_{j=1}^{n} \lbrace k(\boldsymbol{\theta}_{v\,j}^{l},\boldsymbol{\theta}_{v}) \nabla_{\boldsymbol{\theta}_{v\,j}^{l}}\log P(\boldsymbol{\theta}_{v\,j}^{l}|{\bf d})+\nabla_{\boldsymbol{\theta}_{v\,j}^{l}} k(\boldsymbol{\theta}_{v\,j}^{l},\boldsymbol{\theta}_{v}) \rbrace.
\label{eqn:SVGD_vector_improved1}
\end{align}
To obtain $\nabla_{\boldsymbol{\theta}_{v\,j}}\log P(\boldsymbol{\theta}_{v\,j}|{\bf d})$, the calculation of the total derivative is required, because the likelihood function is implicitly dependent on $\boldsymbol{\theta}_{v}$ through $\boldsymbol{\theta}_{T}$, as seen in Equation \ref{eqn:lh_TT}. 
The above calculation can be performed using the Lagrange multiplier method introducing the equality constraints ${\bf r}={\bf 0}$. 
This is a type of the discrete version of the adjoint method \cite{Lewis1985}. 
Here, we define the Lagrange function as the sum of the log likelihood function $J=\log P(\mathbf{T}_{\rm obs} | \boldsymbol{\theta}_{v})$ and a Lagrange multiplier satisfying the above constraints, obtaining
\begin{eqnarray}
J_{\rm L} = J + \boldsymbol{\lambda}^{\top}{\bf r},
\label{eqn:adj_le}
\end{eqnarray}
where $J_{\rm L}$ and $\boldsymbol{\lambda}$ are the Lagrange function and multiplier, respectively. 
The total derivative of $J_{\rm L}$ with respect to $\boldsymbol{\theta}_{v}$ can be calculated using the following chain rule of differentiation:
\begin{align}
\frac{d J_{\rm L}}{d \boldsymbol{\theta}_{v}}
&=\frac{\partial J_{\rm L}}{\partial \boldsymbol{\theta}_{v}} + 
\frac{\partial J_{\rm L}}{\partial \boldsymbol{\theta}_{T}} \frac{d \boldsymbol{\theta}_{T}}{d \boldsymbol{\theta}_{v}} + 
\frac{\partial J_{\rm L}}{\partial \boldsymbol{\lambda}} \frac{d \boldsymbol{\lambda}}{d \boldsymbol{\theta}_{v}}\nonumber\\
&= \boldsymbol{\lambda}^{\top}\frac{\partial {\bf r}}{\partial \boldsymbol{\theta}_{v}} + 
(\frac{\partial J}{\partial \boldsymbol{\theta}_{T}}+\boldsymbol{\lambda}^{\top}\frac{\partial {\bf r}}{\partial \boldsymbol{\theta}_{T}}) \frac{d \boldsymbol{\theta}_{T}}{d \boldsymbol{\theta}_{v}} + 
{\bf r}^{\top} \frac{d \boldsymbol{\lambda}}{d \boldsymbol{\theta}_{v}}. 
\label{eqn:adj_td}
\end{align}
The second and third terms on the right-hand side still include a total derivative. However, the third term vanishes if $f_T$ is trained for $f_v$ and ${\bf r}$ becomes sufficiently small. The second term can also be eliminated by solving the following equation for $\boldsymbol{\lambda}$:
\begin{eqnarray}
\frac{\partial J}{\partial \boldsymbol{\theta}_{T}}+\boldsymbol{\lambda}^{\top}\frac{\partial {\bf r}}{\partial \boldsymbol{\theta}_{T}}={\bf 0}.
\label{eqn:adjoint_equation}
\end{eqnarray}
We approximate the solution of the last equation using the L-BFGS algorithm \cite{Liu1989} using a zero vector as the initial solution. 
Equation \ref{eqn:adjoint_equation} represents a discrete adjoint equation. 
The target total derivative is obtained by evaluating the first term in Equation \ref{eqn:adj_td}, using the solution $\boldsymbol{\lambda}^{*}$ satisfying Equation \ref{eqn:adjoint_equation}:
\begin{eqnarray}
\frac{d J_{\rm L}}{d \boldsymbol{\theta}_{v}}=\boldsymbol{\lambda}^{*\,\top}\frac{\partial {\bf r}}{\partial \boldsymbol{\theta}_{v}}.
\end{eqnarray}
Using this method, the gradient of the logarithm of the posterior PDF can be calculated as follows:
\begin{eqnarray}
\nabla_{\boldsymbol{\theta}_{v}}\log P(\boldsymbol{\theta}_{v}|{\bf d}) 
&=&
\nabla_{\boldsymbol{\theta}_{v}}\log P(\mathbf{T}_{\rm obs} | \boldsymbol{\theta}_{v})+
\nabla_{\boldsymbol{\theta}_{v}}\log P(\boldsymbol{\theta}_{v})\nonumber\\
&=& 
\frac{d J_{\rm L}}{d \boldsymbol{\theta}_{v}}+\nabla_{\boldsymbol{\theta}_{v}}\log P(\boldsymbol{\theta}_{v}).
\end{eqnarray}
The discrete adjoint method is usually applied to regular systems, such as discretized PDEs. 
However, our adjoint equation is over-determined, in which the number of constraints (collocation points) is larger than that of the unknowns (weight parameters). 
Because PINN-based systems of PDEs are usually over-parameterized, the number of the former is significantly larger than the latter. 
For practical applications, considering that $\displaystyle \frac{\partial J}{\partial \boldsymbol{\theta}_{T}}$ is a highly sparse vector due to the over-parametrization, Equation \ref{eqn:adjoint_equation} is approximately satisfied by a moderate number of collocation points. 

After each SVGD update for $\boldsymbol{\theta}_{v}$ for all particles, $\boldsymbol{\theta}_{T}$ is trained for a new $\boldsymbol{\theta}_{v}$ satisfying ${\bf r}={\bf 0}$ 
as follows:
\begin{eqnarray}
\boldsymbol{\theta}_{T}^{*} = \argmin_{\boldsymbol{\theta}_{T}} L(\boldsymbol{\theta}_{T}),
\label{eqn:improved1_argmin}
\end{eqnarray}
where
\begin{eqnarray}
L 
=\frac{1}{N_c} \sum_{i=1}^{N_c} \left( f_{v}(\mathbf{x}_{c}^{(i)},\boldsymbol{\theta}_{v})-\frac{1}{|\nabla f_{T}(\mathbf{x}_{c}^{(i)},\mathbf{x}_{s}^{(i)},\boldsymbol{\theta}_{T})|} \right)^{2}. 
%&+& \frac{1}{N_s} \sum_{i=1}^{N_s}\left( f_{v}(\mathbf{x}_{s}^{(i)}; \boldsymbol{\theta}_{v}) - f_{\tau^{-1}}(\mathbf{x}_{s}^{(i)},\mathbf{x}_{s}^{(i)};\boldsymbol{\theta}_{T}) \right)^{2}. 
\label{eqn:improved1_L}
\end{eqnarray}
The resulting process is equivalent to a PINN-based forward solver of the eikonal equation \cite{Smith2020,Waheed2021PINNeik}.
Algorithm \ref{alg:improved1} summarizes the improved SVGD algorithm in which the target parameter space is reduced to that of $\boldsymbol{\theta}_{v}$ only. 

\subsection{Bayesian estimation of velocity NN with particle-based functional variational inference}
\label{sct:fSVGD}

In the previous section, the parameter space for the Bayesian estimation was reduce to that of $\boldsymbol{\theta}_{v}$. 
However, the NN for $f_v$ is high-dimensional and over-parameterized as well, which result in a posterior PDF with multi-modal features in the weight space that impairs Bayesian estimation. 
To address this, we introduce a function-space SVGD (fSVGD or, more generally, fParVI) \cite{Wang2019} to formulate the Bayesian estimation directly in the function space predicted by the associated NN. 
This reformulation remarkably improves the estimation accuracy of the posterior PDF in BNN by avoiding the direct exploration of a multi-modal PDF over the weight space \cite{Wang2019}. 
We rewrite the equation for the Bayes' theorem as follows:
\begin{eqnarray}
P({\bf v}|\mathbf{d}) &\propto& 
P(\mathbf{T}_{\rm obs} | {\bf v}) P({\bf v}),
\end{eqnarray}
where ${\bf v}$ is evaluated at $N_v$ velocity evaluation points. The evaluation points are randomly chosen, similarly to collocation points:
\begin{eqnarray}
{\bf v} 
&=& \left[ f_v(\mathbf{x}_{v}^{(1)},\boldsymbol{\theta}_{v})\,f_v(\mathbf{x}_{v}^{(2)},\boldsymbol{\theta}_
{v}),\dots,f_v(\mathbf{x}_{v}^{({N_v})},\boldsymbol{\theta}_{v}) \right]^{\top},
\end{eqnarray}
where the associated velocity dataset is defined as 
${\bf X}_v = \{ ({\bf x}_v^{(1)}, {\bf x}_s^{(1)}),({\bf x}_v^{(2)}, {\bf x}_s^{(2)}),\dots,({\bf x}_v^{(N_v)}, {\bf x}_s^{(N_v)}) \}$. 
The source locations are used for the evaluation of the residual vector in the adjoint calculation. 
The update vector in Equation \ref{eqn:SVGD_update_improved1} is modified in fSVGD as follows:
\begin{eqnarray}
\boldsymbol{\phi}\left(\boldsymbol{\theta}_{v\,i}^{l+1}\right)= \frac{\partial {\bf v}_{i}}{\partial \boldsymbol{\theta}_{v\,i}^{l}}^{\top}\boldsymbol{\psi}({\bf v}_{i}^{l}),
\label{eqn:SVGD_vector_improved2}
\end{eqnarray}
where
\begin{eqnarray}
\boldsymbol{\psi}({\bf v})=\frac{1}{n} \sum_{j=1}^{n} \lbrace k({\bf v}_{j}^{l},{\bf v}) \nabla_{{\bf v}_{j}^{l}}\log P({\bf v}_{j}^{l}|\mathbf{d})+\nabla_{{\bf v}_{j}^{l}} k({\bf v}_{j}^{l},{\bf v})\rbrace.
\label{eqn:SVGD_vectorv_improved2}
\end{eqnarray}
The above rule calculates the update vector in the function (velocity) space based on SVGD and then converts it in the weight space using the Jacobian matrix.
As the repulsive force is defined over the function space, the approximation of the posterior PDF does not suffer from the multi-modal feature present in the weight space formulation. 
Furthermore, as the functional prior $P({\bf v})$ is explicitly included, more physically meaningful prior information $P(\boldsymbol{\theta}_{v})$ than that in the weight space can be introduced straightforwardly. 
Although velocity evaluation points can be changed in each epoch, the same points must be used for all the SVGD particles because the point locations define the PDF evaluated by SVGD.
The calculation of $\nabla_{{\bf v}_{j}^{l}}\log P({\bf v}_{j}^{l}|\mathbf{d})$ requires the adjoint formulation $\nabla_{\boldsymbol{\theta}_{v\,j}}\log P(\boldsymbol{\theta}_{v\,j}|{\bf d})$, as in the analysis presented in the previous section. 
In fact, the same formulation in Equations \ref{eqn:adj_le}, \ref{eqn:adj_td} and \ref{eqn:adjoint_equation} is applicable, simply replacing $\boldsymbol{\theta}_{v}$ with ${{\bf v}}$.
These methods overcome the existing limitations in BNN for PINN-based inverse analyses in realistically complex problem setting.
As we focused on the velocity space to perform fSVGD, the final version of the proposed approach was called ``velocity-space SVGD for PINN-based seismic tomography'' (vSVGD-PINN-ST).
Algorithm \ref{alg:improved2} summarizes the steps of vSVGD-PINN-ST.

%% file: sct2.5.tex
\section{Synthetic test in 1D tomographic problem}
\label{sct:linear}

To verify the ability of the proposed method to perform Bayesian estimation, we apply vSVGD-PINN-ST to synthetic 1D tomography tests in a simple problem setting and compare the results with those obtained using the baseline SVGD-PINN-ST and linear travel time tomography, which provides an analytical solution.

\subsection{Linear traveltime tomography}

Linear travel time tomography estimates velocity perturbation from a reference model using a Taylor series expansion. 
We obtain, as a result, a linear inverse problem for the residual travel time. 
When a conjugate pair of the prior and posterior PDF is adopted, such as Gaussian distributions, Bayesian linear regression for linearized tomography provides an analytical solution for the posterior PDF (see Text S1 in  Supporting Material). 
However, linear tomography neglects the dependence of the ray path (i.e., travel time in 1D problems) on velocity perturbations in the reference model. 
Consequently, the method accuracy degrades when the reference model does not offer an accurate approximation of the true one. 

\subsection{1D Synthetic Test}
\label{sct:1DST}

In the 1D synthetic test (1DST), we set a simple true velocity model, with a constant velocity of 1\,km/s in the 1D domain defined by $0 \leq x \leq 1.2\,{\rm km}$, to test the UQ performance (see Fig. \ref{fig:1D_cl0.25}). 
We employ such a simple structure because the focus here is the ability of UQ, not the estimation of velocity.
In this configuration, the travel time is readily obtained. 
Ten points, serving both as receivers and sources, are evenly distributed in two regions defined by the intervals $0.2 \leq x \leq 0.4\,{\rm km}$ and $0.8 \leq x \leq 1\,{\rm km}$ using a 0.05\,km spacing. 
We refer to the five points in each of the above intervals as Group 1 and Group 2, respectively. 
We only consider rays between points within the same group. 
Therefore the number of travel time data points is $5\times4+5\times4=40$.
No ray paths exist in the intervals $0 \leq x \leq 0.2\,{\rm km}$, $0.4 \leq x \leq 0.8\,{\rm km}$, and $1 \leq x \leq 1.2\,{\rm km}$, in which the uncertainty of velocity estimation is expected to be closer to that given by prior. 
${\bf E}_{\rm obs}$ was set assuming an i.i.d. zero-mean Gaussian noise distribution with standard deviation $0.005\,s$, although we did not actually add artificial noise to the travel times, which were calculated analytically between points. 
The prior probability in the velocity space is represented as a stochastic process for Bayesian estimation because $f_v$ is a continuous function of the 1D coordinate $x$ in PINN, evaluated at arbitrary collocation points within the target domain. 
Hence, we use a Gaussian process with the same mean $\mu(x)=1\,{\rm km/s}$ as the true model and kernel function for the Gaussian process defined by: 
\begin{eqnarray}
k_{\rm GP}(x_i, {x}_j)=\sigma_1^2 \exp\left(-\frac{1}{2\sigma_2^2}|{x}_i-{x}_j|^2\right).
\label{eqn:kernel}
\end{eqnarray}
This is an RBF kernel, where $\sigma_1$ and $\sigma_2$ are the standard deviations of the marginal probability and correlation length scale, respectively.
The covariance matrix produced by the RBF kernel is sometimes numerically unstable (i.e., positive definiteness is violated numerically).
To address this instability, we regularized the matrix on the standardized scale following Equation (3) in \cite{Warton2008} with $\lambda=10^{-5}$. 
We confirmed that this regularization does not cause any unintended effects on the results. 
We set $\sigma_1=0.1\,{\rm km/s}$ and adopt three different $\sigma_2$ values, namely, $0.25\,{\rm km}$, $0.15\,{\rm km}$, and $0.075\,{\rm km}$, for comparison. 

In linearized tomography, which we consider to calculate the ground truth of the posterior probability, we set the reference model constant velocity at the true value of 1\,km/s to achieve the best accuracy for the linear approximation. 
We divide the region using a 0.025\,km spacing and estimate 48 unknowns parametrizing the velocity perturbation. 
For these parameters, a prior PDF is generated based on the Gaussian process defined above. 

In vSVGD- and SVGD-PINN-ST, we use fully connected feed-forward neural networks for both $f_{{\tau}^{-1}}$ and $f_{v_{\rm ptb}}$, setting $v_0({\bf x})=1$\,km/s.
The values of $v_{\rm ptb}^{\rm max}$ and $v_{\rm ptb}^{\rm min}$ are set to -0.4 and 0.4\,km/s, giving the upper and lower limits of the velocity prediction $f_{v}$ as 1.4 and 0.6\,km/s, respectively. 
The Swish activation function \cite{Ramachandran2018} is applied in each layer, except in the output one where a linear activation is specified.
We use four hidden layers for both $f_{{\tau}^{-1}}$ and $f_{v_{\rm ptb}}$ with 50 and 10 hidden units, respectively. 
$\boldsymbol{\theta}_{v}$ is initialized using the He's method \cite{He2015}. 
$\boldsymbol{\theta}_{T}$ is trained using the initialized $\boldsymbol{\theta}_{v}$ according to Equations \ref{eqn:improved1_argmin} and \ref{eqn:improved1_L}, before running the algorithms. 
For both algorithms, 256 SVGD particles are employed and the Adam optimizer \cite{Kingma2015} is used to determine $\epsilon_l$. 
The travel time batch data size for ${\bf X}_{T}^{b}$ (see line 2 in Algorithm \ref{alg:improved2}) is set to 40 (i.e., full batch). 
For practical convenience, the coordinate data of the velocity evaluation points ${\bf X}_{v}$ are generated in each iteration by random sampling in the target domain. 
The associated batch data ${\bf X}_{v}^{b}$ are taken as the corresponding ${\bf X}_{v}$ values.
The data size is assigned to 200; 
the number of epochs for each iteration $l$ of vSVGD-PINN-ST is set to 2,000.
Each training session of $\boldsymbol{\theta}_{T}$ (see line 6 in Algorithm \ref{alg:improved2}) is conducted by using a L-BFGS algorithm \cite{Liu1989} for 10 epochs with $N_c = 200$. The collocation points coordinates are generated randomly at each iteration, similarly to the ${\bf X}_{v}$ selection process. 
The initial learning rate of the Adam optimizer is set to $10^{-2}$. 
For the baseline naive SVGD-PINN-ST, introducing a prior probability in the weight space, equivalent to that in the velocity space in Equation \ref{eqn:kernel}, represents a challenging task. 
Hence, we use an i.i.d. zero-mean Gaussian distribution as the prior probability of $\boldsymbol{\theta}=(\boldsymbol{\theta}_{T}\,\boldsymbol{\theta}_{v})^{\top}$ with variance $\sigma_{\theta}^2$.
Considering that a physically meaningful choice of $\sigma_{\theta}$ values is difficult, we test the standard value $\sigma_{\theta}^2=10^0$ (see, for instance, \cite{Yang2021}) and a significantly large one $\sigma_{\theta}^2=10^2$. 
The number of epochs for each iteration $l$ and the initial learning rate of the Adam optimizer are set to 30,000 and $10^{-3}$, respectively.

When the prior probability has a long correlation length ($\sigma_2=0.25\,{\rm km}$), the uncertainty estimation by vSVGD-PINN-ST agrees well with the analytical solution of the linearized tomography (see Fig. \ref{fig:1D_cl0.25} (a) and (b)). 
In the region including the sources and receivers, the standard deviation ($1\sigma$) of the posterior probability is small for both methods. 
This value increases where there is a lack of observations. However, it is always significantly smaller than the standard deviation of the prior probability due to the imposed spatial correlation constraint. 
With an intermediate and a short correlation length ($\sigma_2=0.15\,{\rm km}$ and $0.075\,{\rm km}$), the uncertainty estimation accuracy by vSVGD-PINN-ST slightly decreases, as shown by the slightly overestimated uncertainty at the boundaries (see Fig. \ref{fig:1D_cl0.25} (c)(e) and (d)(f) for linearized tomography and vSVGD-PINN-ST, respectively).
%In particular, for $\sigma_2=0.15\,{\rm km}$ the uncertainty estimated by vSVGD-PINN-ST is close to the analytical solution at both boundaries of the region. However, simulated results deviate from analytical values in the central region.
%For $\sigma_2=0.075\,{\rm km}$, although the estimated standard deviation in the central region is accurate, the curve is smoother than that of the analytical solution on the whole domain. 
%This feature may be also related to. 
The prior probability with a short correlation length assumes the existence of shorter-wavelength components in the target velocity structure. 
The NN architecture adopted in this analysis fails to learn some of these components possibly due to some learning bias, such as the spectral bias \cite{Rahaman2019} that affects the ordinary PINN formulation based on fully connected feed-forward NNs. 
A learning bias may also have slightly compromised the uncertainty estimation accuracy of the experiments with the two lowest $\sigma_2$ values. 
However, we consider the discrepancy in the estimated uncertainty acceptable for practical use at present.
Further advances, in fact, would require improved NN architectures. Alternatively, a different configuration of the loss function may reduce the effect of the learning bias. This aspect is discussed in Section \ref{sct:advantages}, as a future development. 

We examine the relationship between UQ results and two other parameters, namely, the number of particles and epochs, focusing mainly on the experiment with $\sigma_2=0.25\,{\rm km}$.
Surprisingly, even when the number of SVGD particles ($n$) is small (e.g., 32), the estimated posterior mean and standard deviation still capture the basic features of the analytical solution (see Fig. \ref{fig:1D_parameters}(a)). 
This result demonstrates the advantage of using SVGD in the approximation efficiency over Bayesian sequential sampling methods. 
%However, we observe an asymmetry in the shape of the curve of the standard deviation when $n$ is small, showing a imperfect convergence near Group 2. 
%The asymmetry disappears when $n$ is increased. 
Increasing the number of epochs (or iterations) makes the UQ results converge from the side of larger standard deviations (see Fig. \ref{fig:1D_parameters} (b)).
This means that the SVGD update started with sufficiently diverging particles. 
%Interestingly, the standard deviation curves of the first epochs show asymmetric patterns similar to the one seen in the results where a smaller $n$ is used. 
%Another interesting finding is that the same analysis applied to the experiment with $\sigma_2=0.15\,{\rm km}$ shows faster convergence than the previous case (see Fig. S1 in Supporting Material). 
%These results suggest SVGD requires more iterations to converge for the prior probability distributions with longer correlation length scale. 

SVGD-PINN-ST with $\sigma_{\theta}^2=10^0$ completely failed to reproduce the expected spatial variation features of the posterior probability for both of the assumed prior PDFs in the weight space, severely underestimating the standard deviation (Fig. \ref{fig:1D_SVGD}). 
This underestimation does not appear to be an effect of the level of variance in the prior PDF because the result for a larger prior variance, $\sigma_{\theta}^2=10^2$, shows the same tendency (Fig. S2 in Supporting Material). 
% (the one with $\sigma_{\theta}^2=10^0$), showing no correspondence between these features and the locations of sources and receivers (see Fig. \ref{fig:1D_SVGD} and . 
These findings suggest two possible reasons explaining the poor performance of SVGD-PINN-ST in the estimation of the posterior probability.
The first one is that the parameter space for the Bayesian estimation using the naive SVGD approach is too broad and multi-modal;
the second one is that unfavorable effects on the prior probability in the weight space are not physically interpretable. 
Estimated SVGD-PINN-ST mean values are close to the true one. However, they show a larger discrepancy than those calculated by vSVGD-PINN-ST result.

In summary, vSVGD-PINN-ST can estimate the uncertainty in tomographic velocity estimation problems more accurately than the baseline SVGD-PINN-ST method. 
We confirmed that improvements in vSVGD-PINN-ST, which avoid direct Bayesian estimation in high-dimensional and multi-modal weight spaces, are essential to conduct an efficient Bayesian estimation in PINN-based inversion analyses. 
We also found that, even when the spatial correlation length of the velocity is short, vSVGD-PINN-ST returns competitive uncertainty estimates.

%% file: sct3.tex
\section{Synthetic tests in 2D tomographic problems} 
\label{sct:2D}

In this section, we test the applicability of the vSVGD-PINN-ST method to realistic 2D synthetic problems in two scenarios: surface-wave and refraction tomography. 

\subsection{2D Synthetic Test 1: surface-wave tomography} 
\label{sct:circular}

The 2D synthetic test 1 (2DST1) is a synthetic test in 2D surface-wave tomography, in which a low velocity anomaly is surrounded by sources and receivers aligned in a circular arrangement. 
The purpose of this test is to show that the proposed vSVGD-PINN-ST algorithm gives a reasonable estimation of the velocity and its uncertainty, consistent with characteristic ray paths drawn using this circular configuration. 

We use a true velocity model with a homogeneous background set at 2\,km/s, containing a circular low velocity anomaly with 1.2\,km/s at the center. Sixteen receivers are evenly distributed around the anomaly at a radius of 4\,km (see Fig. \ref{fig:circular} (a)). 
Each receiver also serves as a source point, 
thus, the number of travel time data points is $16\times15=240$.
This configuration is inspired by the one in \cite{Galetti2015,Zhang2020Seismic}. However, we use connected background and anomaly regions to comply with the Gaussian prior distribution introduced later, which supports a NN-predicted continuous velocity function. 
We calculate the travel time for each source--receiver pair using the fast sweeping method \cite{Zhao2005} implemented with the ttcrpy python package \cite{Giroux2021}. 
We add independent zero-mean Gaussian noise with standard deviation 0.01\,s to the calculated travel times and use the resulting values as the observation data. 
${\bf E}_{\rm obs}$ is set accordingly, assuming that the error distribution of the travel time observations is known. 
The ray paths drawn using the numerical solution are absent at the center of the anomaly and outside of the sources and receivers (see Fig. \ref{fig:circular} (b)). An accurate Bayesian estimation should infer a larger standard deviation for the estimated velocity in these regions compared to that in the areas where the ray paths go through. 
We adopted a Gaussian process as prior probability, defined by the kernel function in Equation \ref{eqn:kernel} with the 2D L2 norm as argument, mean $\mu({\bf x})=1.75\,{\rm km/s}$, and standard deviations 
$\sigma_1=0.5\,{\rm km/s}$, and $\sigma_2=1.5\,{\rm km}$. 
We regularized the resulting covariance matrix in the same way as in 1DST. 

In vSVGD-PINN-ST, we use fully connected feed-forward neural networks for both $f_{{\tau}^{-1}}$ and $f_{v_{\rm ptb}}$, with $v_0({\bf x})=\mu({\bf x})$.
$v_{\rm ptb}^{\rm max}$ and $v_{\rm ptb}^{\rm min}$ are set to -1.5\,km and 1.5\,km, giving the upper and lower limits of the velocity predicted by $f_{v}$ as 3.25\,km/s and 0.25\,km/s, respectively. 
For both networks, we apply the Swish activation function \cite{Ramachandran2018} in each layer except in the output one where a linear activation is specified.
Six hidden layers are used for both $f_{{\tau}^{-1}}$ and $f_{v_{\rm ptb}}$ with 50 and 20 hidden units, respectively. 
The NN architectures employed here and in the following application were determined based on preliminary experiments, with reference to those in previous studies that performed similar analyses (e.g., \cite{Rasht-Behesht2022}).
The number of SVGD particles employed for this experiment is 512.
The Adam optimizer \cite{Kingma2015} with an initial learning rate of $10^{-2}$ is used to determine $\epsilon_l$. 
Batch sizes for ${\bf X}_{T}^{b}$ and ${\bf X}_{v}^{b}$ (see line 2 in Algorithm \ref{alg:improved2}) are 240 (i.e., full batch) and 400, respectively. 
$\boldsymbol{\theta}_{v}$ is initialized using a prior Gaussian process, obtained by using vSVGD-PINN-ST with zero weight on the travel time observation data. 
$\boldsymbol{\theta}_{T}$ is trained using the initialized $\boldsymbol{\theta}_{v}$ according to Equations \ref{eqn:improved1_argmin} and \ref{eqn:improved1_L} before the vSVGD-PINN-ST algorithm is applied.
The number of epochs for iteration $l$ is set to 800.
Each training step of $\boldsymbol{\theta}_{T}$ (see line 6 in Algorithm \ref{alg:improved2}) is conducted by using the Rectified Adam (RAdam) algorithm \cite{Liu2019} for 1,000 epochs with $N_c = 1,600$ and an initial learning rate of $10^{-4}$, taking the previous SVGD iteration result as the initial guess.
%To prevent the solution from being trapped into local minima due to an ineffective initial guess, the training of $\boldsymbol{\theta}_{T}$ is conducted from scratch for selected epochs (i.e., epoch number 50, 100, 200, and 400). 
The coordinates of the velocity evaluation and collocation points are randomly generated in each epoch within the target domain and ${\bf X}_{v}^{b}$ and ${\bf X}_{c}^{b}$ are set to the corresponding ${\bf X}_{v}$ and ${\bf X}_{c}$ values, respectively.
Use of such random points is expected to prevent the solution from being trapped into local minima, which may be problematic if fixed points are adopted. 

The mean velocity model estimated by vSVGD-PINN-ST agrees well with the true value in the region where ray paths are present (see Fig. \ref{fig:circular_result} (a) and (b)). 
In the regions outside the sources and receivers circumference, the mean is close to the true value probably because of the assumption on spatial correlation.
The posterior PDF standard deviation values in these regions smoothly approach that of the prior Gaussian process (see Fig. \ref{fig:circular_result} (c)).
The estimated values in the central region reflect the absence of ray paths and are, as expected, larger than those in the surrounding areas where the ray paths go through.
However, overall values are smaller than those of the prior uncertainty. This is primarily due to the assumption on the spatial correlation of the prior probability. 
In addition, the absence of ray paths in such a configuration does not allow too large velocity in this region, providing an additional indirect constraint. 
The mean values in the central area are similar to the true ones, suggesting that the true model used is highly consistent with the prior probability. 
To further analyze results, we compared the marginal probability distributions of three points (black marks in Fig. \ref{fig:circular} (b)). 
Histograms in Fig. \ref{fig:circular_hist} show the difference in the uncertainty of the estimated velocity in the regions with and without ray paths (marked by a circle and a square or an inverse triangle, respectively). 
These results suggest that tomographic results obtained by vSVGD-PINN-ST are consistent with the true values and prior probability, and the ray paths distribution for the true velocity model.

To prove that vSVGD-PINN-ST improves UQ accuracy in a realistic problem setting, we conducted the same analysis using the naive method, SVGD-PINN-ST. 
Similarly to the SVGD-PINN-ST analysis in 1DST, we use i.i.d. zero-mean Gaussian prior PDFs as weight parameters (see Text S2 in Supporting Material for other details), taking $\sigma_{\theta}^2=10^{0}$. 
Results show that the velocity models estimated by the SVGD-PINN-ST are not consistent with the true model and ray paths distributions (see Fig. S3 in Supporting Material), reconfirming that SVGD-PINN-ST could not estimate the posterior probability in seismic tomographic problems effectively. 

We also compared the vSVGD-PINN-ST result with a deterministic estimation result obtained by a PINN-based seismic tomographic method \cite{Waheed2021PINNtomo} described in Section \ref{sct:PINNtomo} (Fig. S4 in Supporting Material). 
In the region with ray paths, the estimated velocity model is similar to the mean model by the proposed vSVGD-PINN-ST in Fig. \ref{fig:circular_result} (a). 
Outside the ray coverage, it shows quite a different result with low velocity that is around 1\,km/s. 
Modeling of velocity in such regions without information from ray paths are not controllable unless incorporating the functional prior $P({\bf v})$ as vSVGD-PINN-ST does.

%% file: sct4.tex
\subsection{2D Synthetic Test 2: seismic refraction tomography}
\label{sct:refraction}

In this section, we present the results of the 2D synthetic test 2 (2DST2), in which we consider a refraction tomographic problem where sources and receivers are located on the Earth's surface. We use refracted rays penetrating underground to estimate 2D velocity profiles. 
Compared to 2DST1, this tomographic problem is ill-posed, given the distribution of sources and receivers \cite{Zelt2003}. Furthermore, there exist only a few studies on Bayesian seismic tomography for this type of setting \cite{Ryberg2018}. 
However, in the following, we show that our vSVGD-PINN-ST algorithm is suitable for this type of tomographic problem as well.

We set a relatively simple true velocity model with an almost depth-dependent structure in a domain ranging from 0 to 30 and from 0 to 200\,km in depth and horizontal distance, respectively. A velocity bump is included in the region between 70 and 130\,km in the horizontal axis.
We consider 20 sources and 96 receivers distributed with a uniform spacing between 5 and 195\,km in the horizontal distance on the surface of the model (see Fig. \ref{fig:refraction} (a)).
Therefore, the total number of travel time data is $20\times96=1920$.
As in 2DST1, we create synthetic data by calculating the travel time for each source-receiver pair using the fast sweeping method and add the i.i.d. zero-mean Gaussian noise with a standard deviation of 0.05\,s. 
${\bf E}_{\rm obs}$ is set accordingly, assuming that the error distribution of traveltime observations is known. 
The ray paths drawn by using the numerical solution are densely distributed in a shallow portion of the target region (see Fig. \ref{fig:refraction} (b)). 
For the prior probability, we set a Gaussian process with an RBF kernel, similarly to the previous tests. 
In refraction tomography, we assume that velocity increases with increasing depth. 
Therefore, we adopted a mean velocity model $\mu({\bf x})$ that is linearly depth-dependent (see Fig. S5 in Supporting Material). 
Under such an assumption, the correlation length in the horizontal and vertical directions are likely to have significant differences. 
To reflect this prior information, we redefine the RBF kernel for the Gaussian process as follows:
\begin{eqnarray}
k_{\rm GP}({\bf x}_i, {\bf x}_j)=\sigma_1^2 \exp\left[-\frac{1}{2}\left(\frac{(x_i-x_j)^2}{\sigma_x^2}+\frac{(z_i-z_j)^2}{\sigma_z^2}\right)\right]
\label{eqn:kernel_2D}
\end{eqnarray}
where ${\bf x} = (x\,z)^{\top}$, and $\sigma_x$ and $\sigma_z$ are the correlation lengths in the horizontal and vertical direction, respectively. 
We set $\sigma_1=1\,{\rm km/s}$, $\sigma_x=35\,{\rm km}$, and $\sigma_z=10\,{\rm km}$.
We regularized the resulting covariance matrix in the same way as in 1DST. 

We use the same fully connected feed-forward neural networks as in the previous experiment, with six hidden layers for both $f_{{\tau}^{-1}}$ and $f_{v_{\rm ptb}}$, including 50 and 20 hidden units, respectively.
We set $v_0({\bf x})=\mu({\bf x})$ as the mean velocity of the prior Gaussian process introduced in the following. 
$v_{\rm ptb}^{\rm max}$ and $v_{\rm ptb}^{\rm min}$ are set to -3\,km and 3\,km, respectively, giving the depth-dependent upper and lower limits of the velocity predicted by $f_{v}$.
The Adam optimizer with an initial learning rate of $10^{-3}$ is used to determine $\epsilon_l$ in both algorithms. 
Weight parameters are initialized in the same way as in 2DST1. 
The batch sizes of ${\bf X}_{T}^{b}$ and ${\bf X}_{v}^{b}$ (see line 2 in Algorithm \ref{alg:improved2}) are 480 and 400, respectively.
The number of epochs for each iteration $l$ is set to 200.
Each training session of $\boldsymbol{\theta}_{T}$ (see line 6 in Algorithm \ref{alg:improved2})is performed by the RAdam algorithm \cite{Liu2019} for 1,000 epochs with $N_c = 1,600$ and an initial learning rate of $10^{-4}$.
The initial guess was given in the same way as in 2DST1. 
To prevent the solution from being trapped into local minima due to an ineffective initial guess, the training of $\boldsymbol{\theta}_{T}$ is conducted from scratch for selected epochs (i.e., epoch number 50 and 100).
%, except for the epochs 50 and 100 that were trained from scratch. 
The coordinates of the velocity evaluation and collocation points were randomly generated in each epoch within the target domain, assigning to ${\bf X}_{v}^{b}$ and ${\bf X}_{c}^{b}$ the corresponding values in ${\bf X}_{v}$ and ${\bf X}_{c}$, respectively.

The mean velocity model estimated by vSVGD-PINN-ST agrees well with the true value in the region with a high density of ray paths (see Fig. \ref{fig:refraction_result} (a) and (b)). 
Fig. \ref{fig:refraction_result} (c) shows that the standard deviation is generally small in the same regions, with some irregular increases marked in light blue that may reflect complex ray-path patterns and heterogeneous velocity structure. 
In the bottom and side region not reached by ray paths, it smoothly increases to around 1\,km/s, the standard deviation of the prior probability.
The results of the experiment are consistent with the true values and the prior probability, proving that our vSVGD-PINN-ST algorithm can estimate the posterior probability successfully, even in highly ill-posed problem setting. 

To further analyze the results, we compared the estimated mean model, true model, prior mean, and marginal probability distributions along the two lines marked in Fig. \ref{fig:refraction} (b) (see Fig. \ref{fig:refraction_hist}). 
In both profiles, the true velocity model agrees well with the estimated mean or, at least, its values are included within a high frequency region in the shallow portion. 
The mean and true velocity begin to grow apart at about 15 and 20\,km of depth in the 20 (dashed line) and 100\,km (dash-dot line) horizontal lines, respectively. We hypothesize that this is due to the reduction in the number of ray paths as the depth increases.
At 20\,km in the horizontal distance, the uncertainty is small even in the depth between 15 and 20\,km that is out of the ray paths coverage, probably due to the spatial correlation assumption in the prior probability. 
Below these depths, the frequency color maps show a broad distribution, with increasing uncertainty due to the lack of ray paths. 
We expected the estimated mean model to agree well with the prior one in this depth, since information is not obtained directly from data. However, the former does not approach the latter as the depth increases, it grows apart from it. 
As we hypothesize in Section \ref{sct:1DST}, we attribute this finding to some learning bias affecting our NN architectures, such as the spectral bias \cite{Rahaman2019}. 
A possible way to address this issue is discussed as a further development in Section \ref{sct:advantages}.

%% file: sct5.tex
\section{Discussion}
\label{sct:discussion}

\subsection{Advantages and future developments of PINN-based Bayesian seismic tomography}
\label{sct:advantages}

The test problem setup for seismic refraction tomography in 2DST2 is different from the two previous ones, presenting the same observation geometry as that of real subsurface structural exploration.
The successful performances of our method in this test confirm its applicability to actual observational data in geophysical exploration and subsurface structural studies in seismogenic zones. 
Existing seismic tomographic methods used in these research domains consider an arbitrary initial model to calculate the theoretical travel time and, subsequently, they update the velocity model using an iterative method to minimize the residual between the theoretical and observed travel time. 
The final velocity model is obtained when the residual becomes sufficiently small.
A Monte Carlo (MC) analysis with initial model randomization is used to evaluate the uncertainty and reliability of the solution obtained using such tomographic methods \cite{Korenaga2000,Kodaira2014}.
The idea behind the MC analysis is that the main source of uncertainty in the estimation results of traditional tomographic methods lies in the choice of the initial model.
The introduction of Bayesian estimation in the proposed method mitigates the dependence of the estimation upon the initial value. The uncertainty evaluation reflects the quality and quantity of the observation data instead.
In conventional tomography methods, model parameterization (quantity and arrangement selection) and regularization (e.g., smoothing parameters selection) are often determined subjectively.
In this study, model parameterization is performed controlling the continuous functions represented by NNs, using the prior probabilities introduced by Bayes' theorem in the form of stochastic processes.
The selection of regularization parameters is performed choosing the ones that characterize prior probabilities in Bayesian statistics, such as the correlation distance in the stochastic process adopted in our method.
Although this aspect is not considered in this study, the most suitable parameters can be objectively determined within the framework of Bayesian statistics (e.g., using hierarchical or empirical Bayesian frameworks, see \cite{Bishop2006}). This aspect represents an important future development.

PINN-based Bayesian seismic tomography benefits from the general advantages of PINN methods for the solution of PDEs and deterministic inversion problems (see \cite{Karniadakis2021}, for instance).
The PINN-based approach we used does not require, in fact, a mesh or grid for numerical calculation because the continuous functions defined by NNs are differentiable and automatically generated in the domain of interest. 
Moreover, neural networks, automatic differentiation, and optimization algorithms can be introduced easily using existing deep-learning frameworks, such as Pytorch and TensorFlow. 
These functions also contributed to the efficient implementation of the SVGD algorithm and the adjoint method required in vSVGD-PINN-ST. 
The extension of this approach to higher dimensional domains (i.e., from 1D to 2D and from 2D to 3D) is easier than in ordinary numerical simulations since the mesh-free framework reduces the dependency of the algorithm on the dimension of the target problem.
In fact, our Pytorch code for vSVGD-PINN-ST required the modification of less than 100 lines of code to switch from 1D to 2D analyses.

The traditional PINN formulation with fully connected feed-forward NNs shows poor performance when the target functions include high-frequency or multi-scale features \cite{Wang2021}, representing higher structural complexity. 
This is due to the NN spectral bias \cite{Rahaman2019}, which may be also responsible for the slight degradation of the estimation accuracy when the correlation length is small in 1DST and the discrepancy between the estimated and prior mean velocity at the bottom of the target domain in 2DST.
Further improvements, such as the introduction of adaptive activation functions \cite{Jagtap2020Adaptive}, Fourier features \cite{Tancik2020,Wang2021}, domain decomposition techniques \cite{Jagtap2021Extended} and loss functions that account for physical causality \cite{Wang2022Causality}, may be important to address the challenge of learning high-frequency components or multi-scale features.
Introduction of dense and heterogeneous collocation points \cite{Mao2020} brought by computer power may be also important for addressing the issue.
We expect that a combination of these improvements will improve the UQ of our PINN-based seismic tomography method, resulting in increasingly realistic velocity structures. 

All calculations performed for vSVGD-PINN-ST were accelerated using full parallelization for each SVGD particle. 
For instance, performing 2DST1 took 27.0 hours, using 512 CPU cores (64-core AMD EPYC 7742 $\times$ 8 in Earth Simulator 4, made available by the Japan Agency for Marine-Earth Science and Technology (JAMSTEC)) assigned to each one of the 512 particles. 
The 2DST1 experiment mimics the setting of the synthetic test by Zhang \& Curtis in \cite{Zhang2020Seismic}, 
where the authors analyzed the computational cost required to execute the several Bayesian seismic tomography methods compared in their study.
They found that analyses based on the Metropolis-Hastings Markov Chain Monte Carlo (MCMC) \cite{Metropolis1953,Hastings1970} and reversible jump MCMC (rj-MCMC) method \cite{Green1995,Bodin2009} required 80.05 and 17.1 calculation hours, respectively, using six CPUs. 
Further parallelization of these sampling algorithms is difficult due to their sequential features. 
These considerations suggest that vSVGD-PINN-ST is competitive with some existing methods in terms of time-to-solution because of the high parallelism of the algorithm (note that Zhang \& Curtis proposed highly efficient variational inference methods \cite{Zhang2020Seismic}). 
Furthermore, the estimation using our PINN-based method provides mesh-free continuous velocity models (such as those of individual SVGD particles shown in Fig. S6 in Supporting Material), whereas those obtained in \cite{Zhang2020Seismic} are parametrized with a relatively coarse spatial grid, which reduces computation cost.
Incorporating finer grids will impose significantly larger cost on these other methods. 
Most of the vSVGD-PINN-ST computational time is required by the PINN-based solution of the eikonal equation in each SVGD update (in line 6 of Algorithm \ref{alg:improved2}). 
Hyper parameters affecting this calculation time, such as the number of hidden layers in the NNs, hidden units in each NN layer, epochs, and collocation points, were not optimized in this study. 
To improve the SVGD efficiency, for instance, reducing the number of particles and iterations, and using improved kernels \cite{Wang2019matrix-valued,Ai2022} and second-order methods \cite{Detommaso2018,Zhu2020} will be considered for future developments. 
For large-scale problems, requiring increasing numbers of NNs hidden layers and units, GPUs efficient acceleration might prove more advantageous for our PINN-based method compared with other ones based on ordinary numerical calculation.

To verify the accuracy of the PINN-based travel time calculation in vSVGD-PINN-ST,
we compared the final travel time predicted by $f_T$ with the result of the fast sweeping method, which uses the final velocity predicted by $f_v$. 
Such comparison with reference solutions obtained from ordinary numerical simulation methods is currently the only available method to check the convergence to the true solution. 
Further theoretical studies are required to understand convergence properties of PINN-based solutions \cite{Karniadakis2021}.

\subsection{Prior probability}

Compared to the 2D surface wave tomography targeted in 2DST1, 2D seismic refraction tomography in 2DST2 is more likely to lead to a highly ill-posed inverse problem due to the distribution of sources and receivers \cite{Zelt2003,Ryberg2018}. 
Leveraging prior information to provide proper constraints is crucial in this type of seismic tomography. 
In 2DST2, we introduced a mean function and correlation length dependent on depth and direction, respectively, in the Gaussian process with a RBF kernel used as the prior probability. 
This choice is based on the solid Earth science knowledge that seismic velocity structures are nearly horizontally stratified. 
The values of the parameters used in the kernel of the prior Gaussian process are reviewed in Table \ref{tab:correlation_length}. 
In our vSVGD-PINN-ST, such prior constraints can be incorporated directly because the Bayesian inference is performed in velocity space. 
In previous studies on Bayesian PINN, the estimation was performed in weight space using a simple prior probability, such as the i.i.d. Gaussian distribution, for the weight parameters \cite{Yang2021,Psaros2023}.
To understand how such a simple prior probability defined in weight space behaves in the physical space, we examine the corresponding Gaussian process in the function space to this prior. 
We draw 1,000 random samples of the weight parameter set of the NN predicting velocity used in 2DST2, using the i.i.d. Gaussian prior distribution, and we generate 1,000 corresponding predicted velocity structures. 
Subsequently, we find the best fitting $\sigma_1$, $\sigma_x$ and $\sigma_z$ values in Equation \ref{eqn:kernel_2D} for the 1,000 velocity models (see Text S3 in Supporting Material for further details).
In Table \ref{tab:correlation_length}, two examples of the estimated parameter sets for the kernel function are presented.
For instance, when the standard deviations of the i.i.d. Gaussian distributions is set to $\sigma_{\theta}=10^{-0.6}$, the marginal standard deviation ($\sigma_1$) has a similar value to that of the prior distribution in 2DST. 
However, the estimated horizontal correlation length ($\sigma_x$) is larger than the horizontal domain size, which is physically not appropriate. 
In contrast, when $\sigma_{\theta}=10^{-0.4}$, although the correlation length appears to be within an acceptable range, the marginal standard deviation ($\sigma_1$) is so large that the constraint from the depth-dependent mean function becomes too weak. 
These two examples demonstrate that introducing a physically interpretable prior probability is not an easy task when Bayesian PINN is performed in the weight space and a simple function such as i.i.d. Gaussian is adopted. A learning algorithm has been proposed in \cite{Tran2022} to obtain the prior probability in the weight space from a target one defined in the function space. However, the high computational complexity of the learning algorithm is inevitably problematic when the problem size becomes large. 
Introducing SVGD in the function space for Bayesian PINN is an effective solution not only because it overcomes the multi-modality issues in the weight space, but also because it results in a physically interpretable Bayesian inference. 

When comparing different methods, we should always consider that a different parametrization leads to a different prior probability, resulting in a consequent discrepancy in the posterior probability.
In \cite{Zhang2020Seismic}, for instance, the results obtained with methods based on adaptive parametrization (i.e., the rj-MCMC method \cite{Green1995,Bodin2009}) were significantly different from those based on a fixed parametrization. 
Similarly, even if we had considered the exact same problem setting as the one in \cite{Zhang2020Seismic}, performing a meaningful comparison would have been difficult because the seismic velocity obtained by the PINN-based tomography is also adaptively parameterized.
Therefore, comparison of Bayesian seismic tomography methods in standardized problem settings with equivalent prior probabilities represents an important aspect to be considered for future developments. 

Previous studies claim that the PINN-based approach can effectively solve ill-posed inverse problems without introducing prior knowledge for the target parameters estimation \cite{Waheed2021PINNtomo,Rasht-Behesht2022}. 
From the Bayesian viewpoint, a PINN-based inverse analysis without prior constraints can be interpreted as a maximum a posteriori (MAP) estimation of the weight parameters, incorporating a prior PDF with uniform distribution of a wide value range (e.g., an improper flat prior). 
If we focus on the space of the function predicted by a NN, such a prior PDF is not a ``non-informative prior'' any more, because it imposes implicit prior information in the function space, obtained by a nonlinear transformation from the weight space (see \cite{Bishop2006} for uniform distributions defined by a nonlinear variable transformation from a different feature space).
As a result, a MAP solution obtained without introducing explicit prior constraints may include effects from implicit prior information that are nonnegligible. 
As proposed in this study, explicitly providing physically interpretable prior information defined in the function space helps in addressing this issue hidden in ordinary PINN-based inverse analyses. 

\subsection{Application of vSVGD-PINN-ST to other PINN-based inversion problems}

The proposed vSVGD-PINN-ST algorithm can be applied to general PINN-based inversion problems that incorporate two NNs, one predicting the solution of the governing equation and the other its parameters (e.g., the full waveform inversion \cite{Rasht-Behesht2022}). 
Some of the previous studies on Bayesian PINN targeted simple problems in which the solution of the governing equation predicted by a single NN is the only target of Bayesian estimation (e.g., \cite{Sun2020,Linka2022}). 
The proposed approach can also be used for this type of problems by introducing a simplified method that we call ``function-space SVGD for PINN'' (fSVGD-PINN). fSVGD-PINN can be derived from vSVGD-PINN-ST simply incorporating fSVGD in PINN as described in Section \ref{sct:fSVGD}, removing the procedure used to separate one of the two NNs from the Bayesian estimation described in Section \ref{sct:adjoint}.

%% file: sct6.tex
\section{Conclusion}
\label{sct:conclusion}

In this study, we developed the vSVGD-PINN-ST algorithm, which performs PINN-based Bayesian seismic tomography using SVGD, the best-known particle-based variational inference method, applied only in the velocity space and enhanced with several mathematical and numerical techniques. 
The vSVGD-PINN-ST performance was tested in one- and two-dimensional Bayesian seismic tomography synthetic tests. Such problems cannot be handled by naive baseline algorithms that perform SVGD in the weight space of the component NNs predicting velocity and travel time. 
Results show that our method not only allows for accurate UQ but it can also incorporate physically-interpretable prior probability defined in the velocity (function) space, overcoming existing limitations of traditional BNN approaches based on Bayesian estimation in the weight space.
To the authors' best knowledge, this is the first success in PINN-based Bayesian seismic tomography with practical estimation accuracy. 
The success of the last synthetic test adopting a realistic observation geometry, similar to a subsurface structural exploration, suggest that our method can be applied to actual observational data in geophysical exploration and subsurface structural studies.
Finally, the proposed method offers a new fundamental Bayesian approach that can be applied to inverse problems sharing the same formulation, in geoscience and other fields, leveraging on the flexibility and extendibility of PINN.

%% file: ack.tex
%\section*{Acknowledgement}

We thank two anonymous reviewers for their careful reviews and constructive comments.
Comments from Dr. Tatsu Kuwatani were valuable for designing the 1D synthetic test. 
This research was supported by JSPS KAKENHI Grant Number 21K14024. 
Computational resources of the Earth Simulator 4 provided by JAMSTEC was used.

%% file: algo.tex
\begin{algorithm}[]
\caption{SVGD-PINN-ST, a naive approach based on SVGD in weight space.}
\label{alg:naive}
\begin{algorithmic}[1]
\renewcommand{\algorithmicrequire}{\textbf{Input:}}
\renewcommand{\algorithmicensure}{\textbf{Output:}}
\REQUIRE A set of initial particles $\lbrace \boldsymbol{\theta}^{0} \rbrace ^n_{i=1}$, where $\boldsymbol{\theta}=(\boldsymbol{\theta}_{T}\,\boldsymbol{\theta}_{v})^{\top}$.
\ENSURE  A set of particles  $\lbrace \boldsymbol{\theta} \rbrace ^n_{i=1}$, which approximates the target distribution $P(\boldsymbol{\theta}|\mathbf{d})$. 
\FOR {iteration $l$}
 \STATE Sample a mini batch ${\bf X}_{T}^{b}$ and ${\bf X}_{c}^{b}$ from training set ${\bf X}_{T}$ and ${\bf X}_{c}$.
 \STATE For each $i \in [n] $, calculate the SVGD update vector for ${\bf X}_{T}^{b}$ and ${\bf X}_{c}^{b}$ according to Equation \ref{eqn:SVGD_vector}.
 \STATE For each $i \in [n] $, calculate $\boldsymbol{\theta}_i^{l+1}$ according to Equation \ref{eqn:SVGD_update}.
 \STATE Set $l \leftarrow l+1$
\ENDFOR
\end{algorithmic}
\end{algorithm}

\begin{algorithm}[]
\caption{An updated approach based on SVGD only in the weight space of velocity NN.}
\label{alg:improved1}
\begin{algorithmic}[1]
\renewcommand{\algorithmicrequire}{\textbf{Input:}}
\renewcommand{\algorithmicensure}{\textbf{Output:}}
\REQUIRE A set of initial particles $\lbrace \boldsymbol{\theta}_v^{0} \rbrace ^n_{i=1}$ and $\lbrace \boldsymbol{\theta}_T^{0} \rbrace ^n_{i=1}$ that are initially trained for $\lbrace \boldsymbol{\theta}_v^{0} \rbrace ^n_{i=1}$. 
\ENSURE  A set of particles  $\lbrace \boldsymbol{\theta}_v \rbrace ^n_{i=1}$, which approximates the target distribution $P(\boldsymbol{\theta}_{v}|\mathbf{d})$. 
\FOR {iteration $l$}
 \STATE Sample a mini batch ${\bf X}_{T}^{b}$ and  ${\bf X}_{c}^{b}$ from training set  ${\bf X}_{T}$ and  ${\bf X}_{c}$.
 \STATE For each $i \in [n] $, calculate the SVGD update vector for ${\bf X}_{T}^{b}$ according to Equation \ref{eqn:SVGD_vector_improved1}.
 \STATE For each $i \in [n] $, calculate $\boldsymbol{\theta}_{v\,i}^{l+1}$ according to Equation \ref{eqn:SVGD_update_improved1}.
 \STATE For each $i \in [n] $, update $\boldsymbol{\theta}_{T\,i}^{l+1}$ for ${\bf X}_{c}^{b}$ according to Equation \ref{eqn:improved1_argmin} and \ref{eqn:improved1_L}.
 \STATE Set $l \leftarrow l+1$
\ENDFOR
\end{algorithmic}
\end{algorithm}

\begin{algorithm}[]
\caption{VSVGD-PINN-ST, the final version of the algorithm based on velocity-space SVGD.}
\label{alg:improved2}
\begin{algorithmic}[1]
\renewcommand{\algorithmicrequire}{\textbf{Input:}}
\renewcommand{\algorithmicensure}{\textbf{Output:}}
\REQUIRE A set of initial particles $\lbrace \boldsymbol{\theta}_v^{0} \rbrace ^n_{i=1}$ and $\lbrace \boldsymbol{\theta}_T^{0} \rbrace ^n_{i=1}$ that are initially trained for $\lbrace \boldsymbol{\theta}_v^{0} \rbrace ^n_{i=1}$. 
\ENSURE  A set of particles $\lbrace \boldsymbol{\theta}_v \rbrace ^n_{i=1}$, such that $f({\bf x}; \boldsymbol{\theta}_{v\,i})$ approximates the target distribution $P(v({\bf x})|\mathbf{d})$. 
\FOR {iteration $l$}
 \STATE Sample a mini batch ${\bf X}_{T}^{b}$, ${\bf X}_{c}^{b}$ and ${\bf X}_{v}^{b}$ from training set ${\bf X}_{T}$, ${\bf X}_{c}$ and ${\bf X}_{v}$.
 \STATE For each $i \in [n] $, calculate the SVGD update vector for ${\bf X}_{T}^{b}$ and ${\bf X}_{v}^{b}$ in the velocity space according to Equation \ref{eqn:SVGD_vectorv_improved2}.
 \STATE For each $i \in [n] $, calculate the SVGD update vector in the weight space according to Equation \ref{eqn:SVGD_vector_improved2}.
 \STATE For each $i \in [n] $, calculate $\boldsymbol{\theta}_{v\,i}^{l+1}$ according to Equation \ref{eqn:SVGD_update_improved1}.
 \STATE For each $i \in [n] $, update $\boldsymbol{\theta}_{T\,i}^{l+1}$ for ${\bf X}_{c}^{b}$ according to Equation \ref{eqn:improved1_argmin} and \ref{eqn:improved1_L}.
\ENDFOR
\end{algorithmic}
\end{algorithm}

%% file: fig.tex
\begin{figure*}
\begin{center}
\begin{small}
\includegraphics[clip, width=12cm, bb = 0 0 970 763]{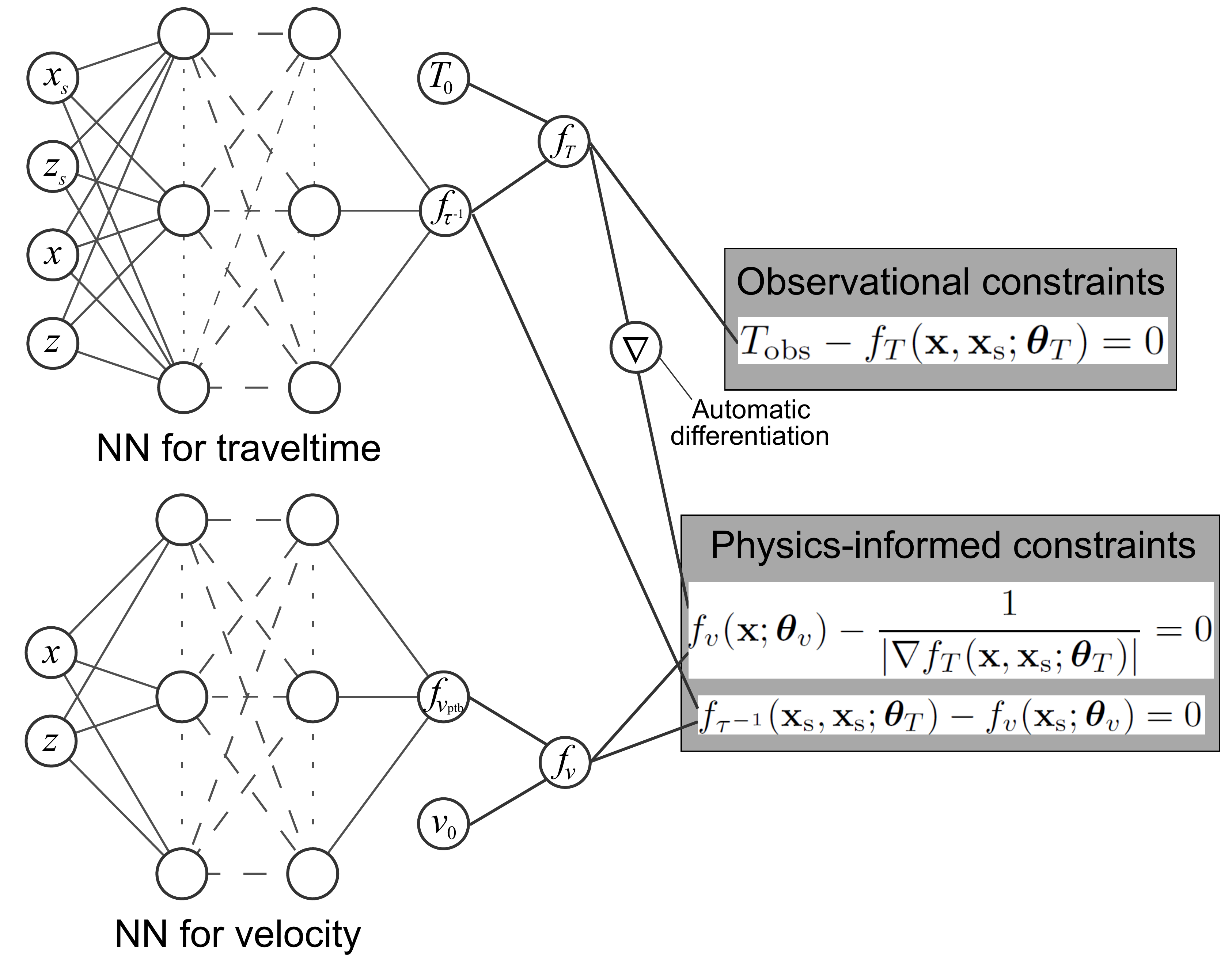}
\end{small}
\end{center}
\caption{Neural network formulation adopted in this study and a schematic view of PINN-based (deterministic) seismic tomography. }
\label{fig:NN_shematics}
\end{figure*}

\begin{figure*}
\begin{center}
\begin{small}
\begin{tabular}{cc}
\includegraphics[clip, width=8cm, bb = 27 42 403 403]{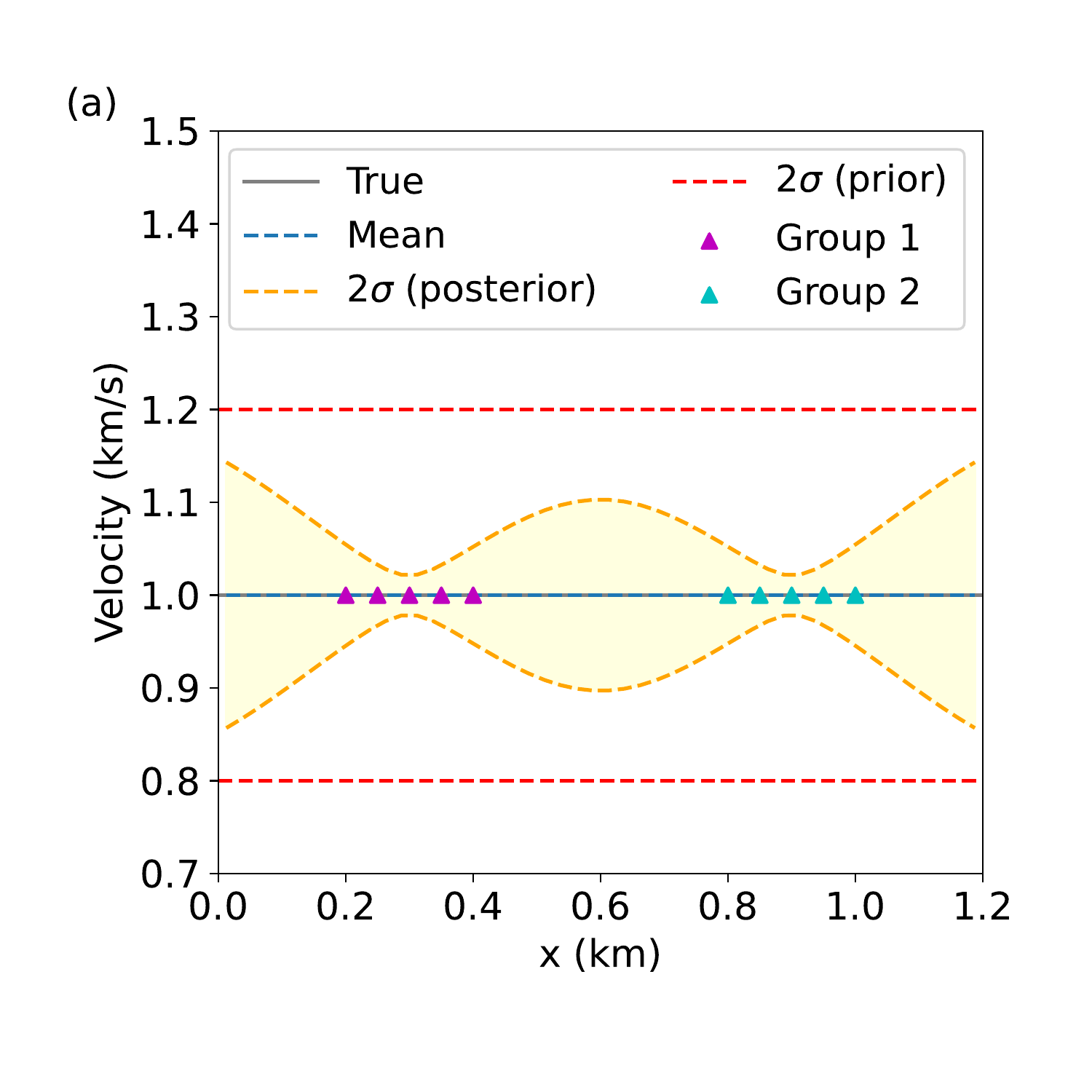}&
\includegraphics[clip, width=8cm, bb = 27 42 403 403]{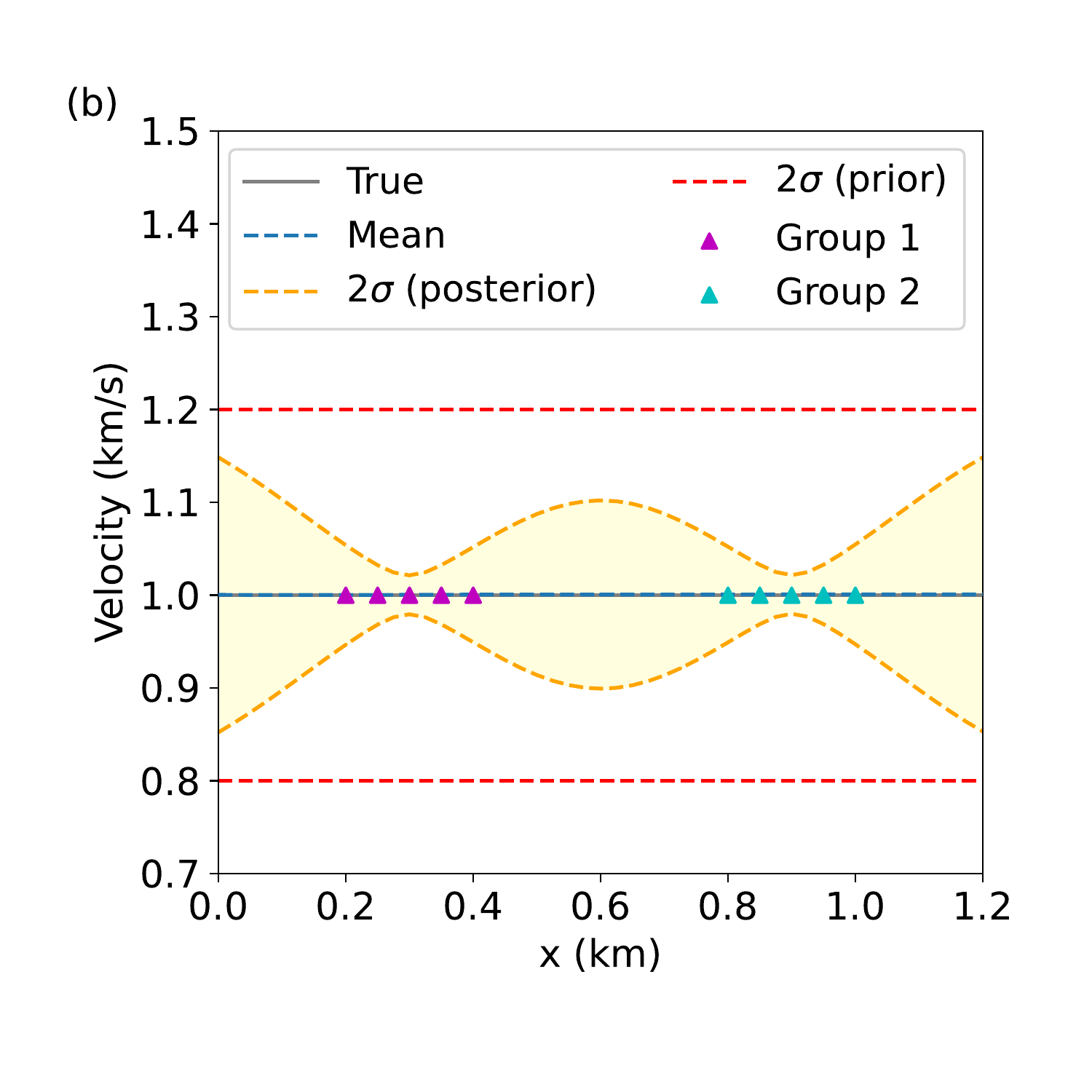}\\
\includegraphics[clip, width=8cm, bb = 27 42 403 403]{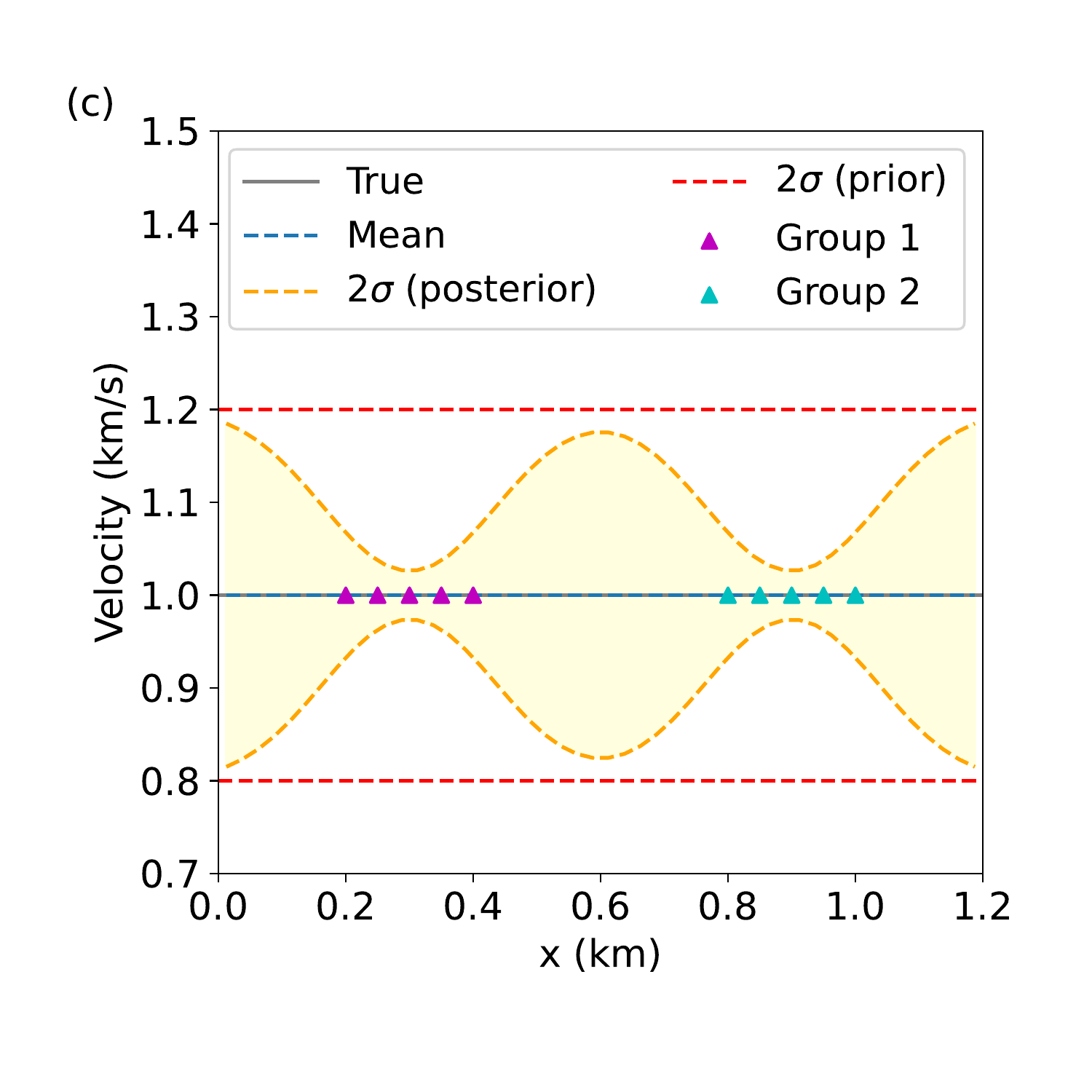}&
\includegraphics[clip, width=8cm, bb = 27 42 403 403]{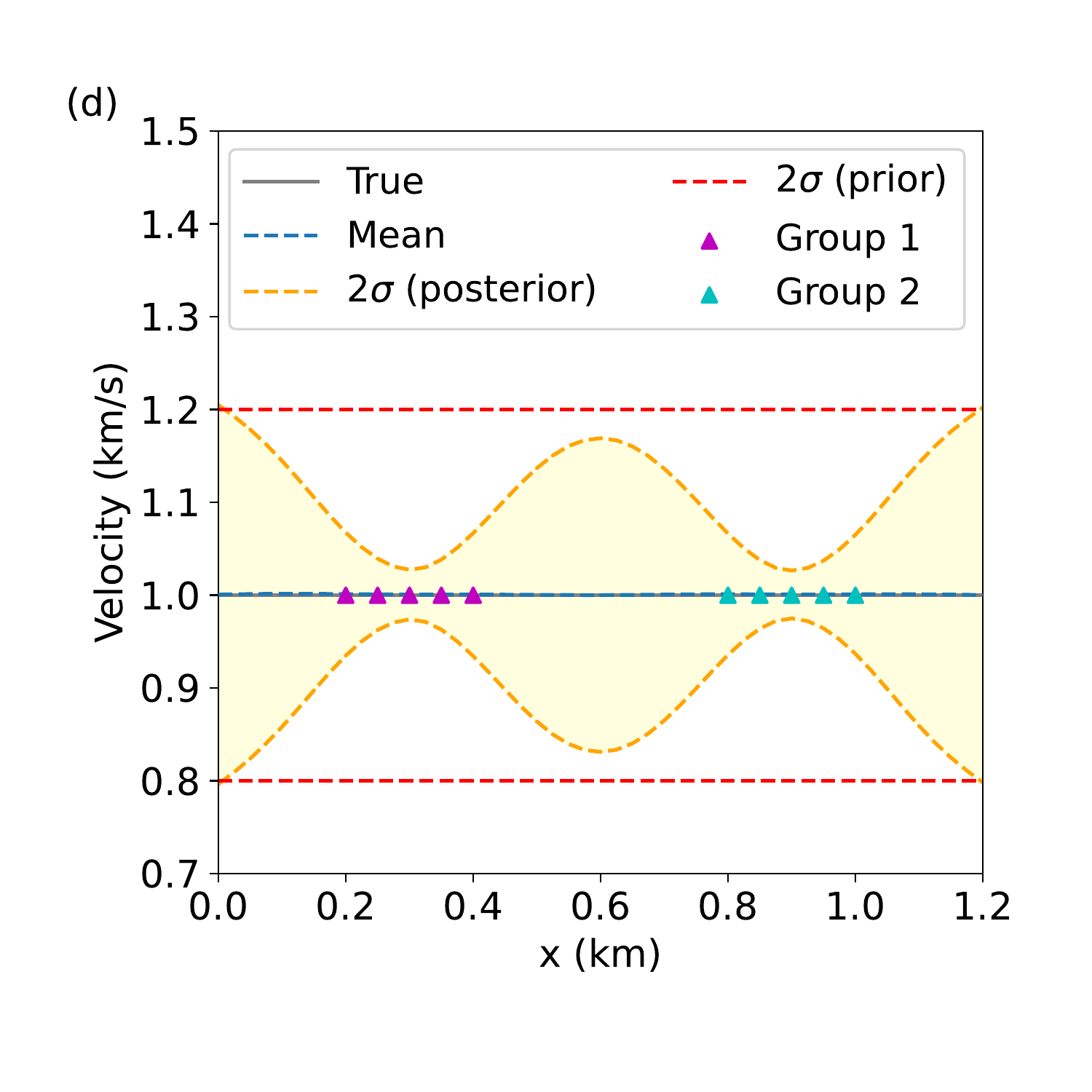}\\
\includegraphics[clip, width=8cm, bb = 27 42 403 403]{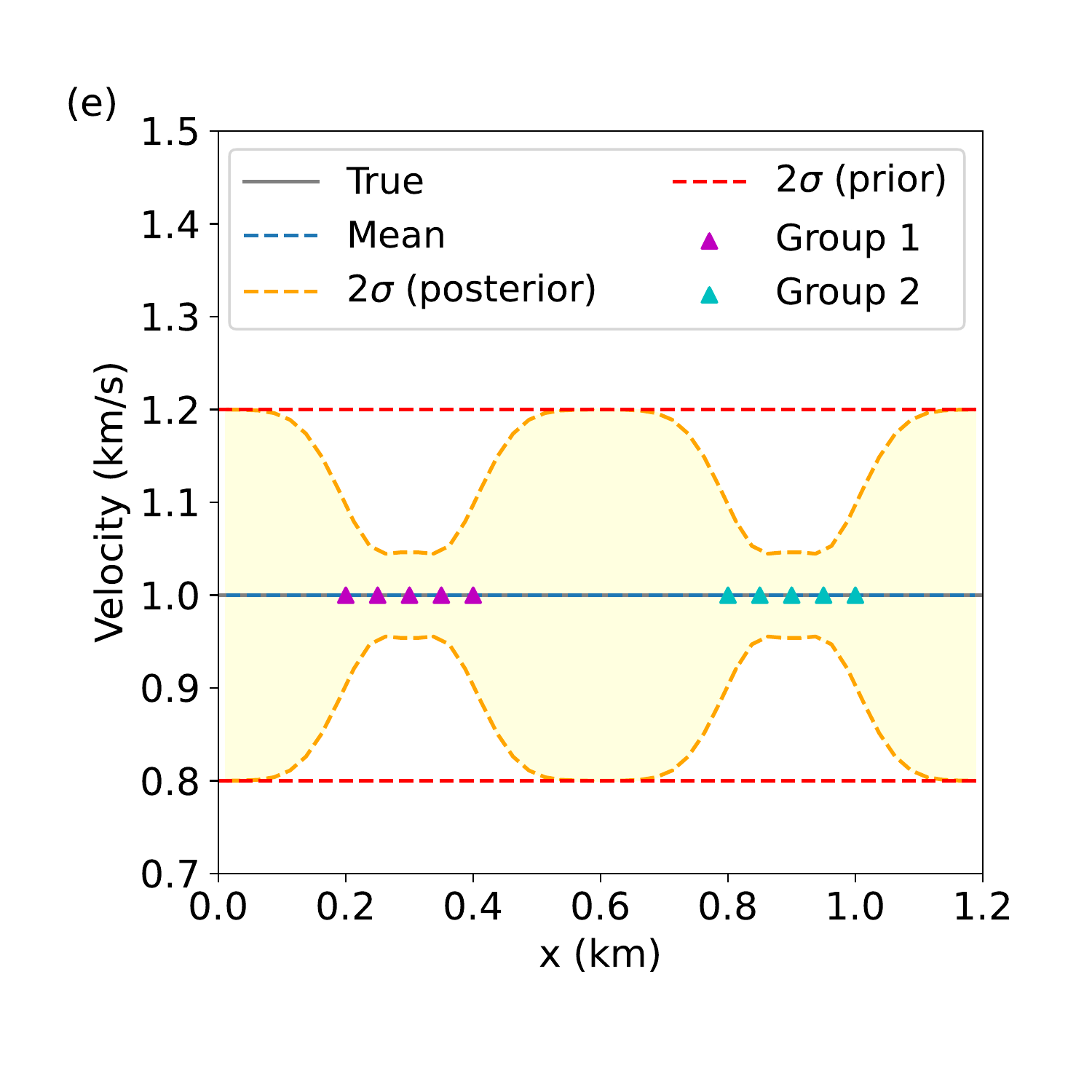}&
\includegraphics[clip, width=8cm, bb = 27 42 403 403]{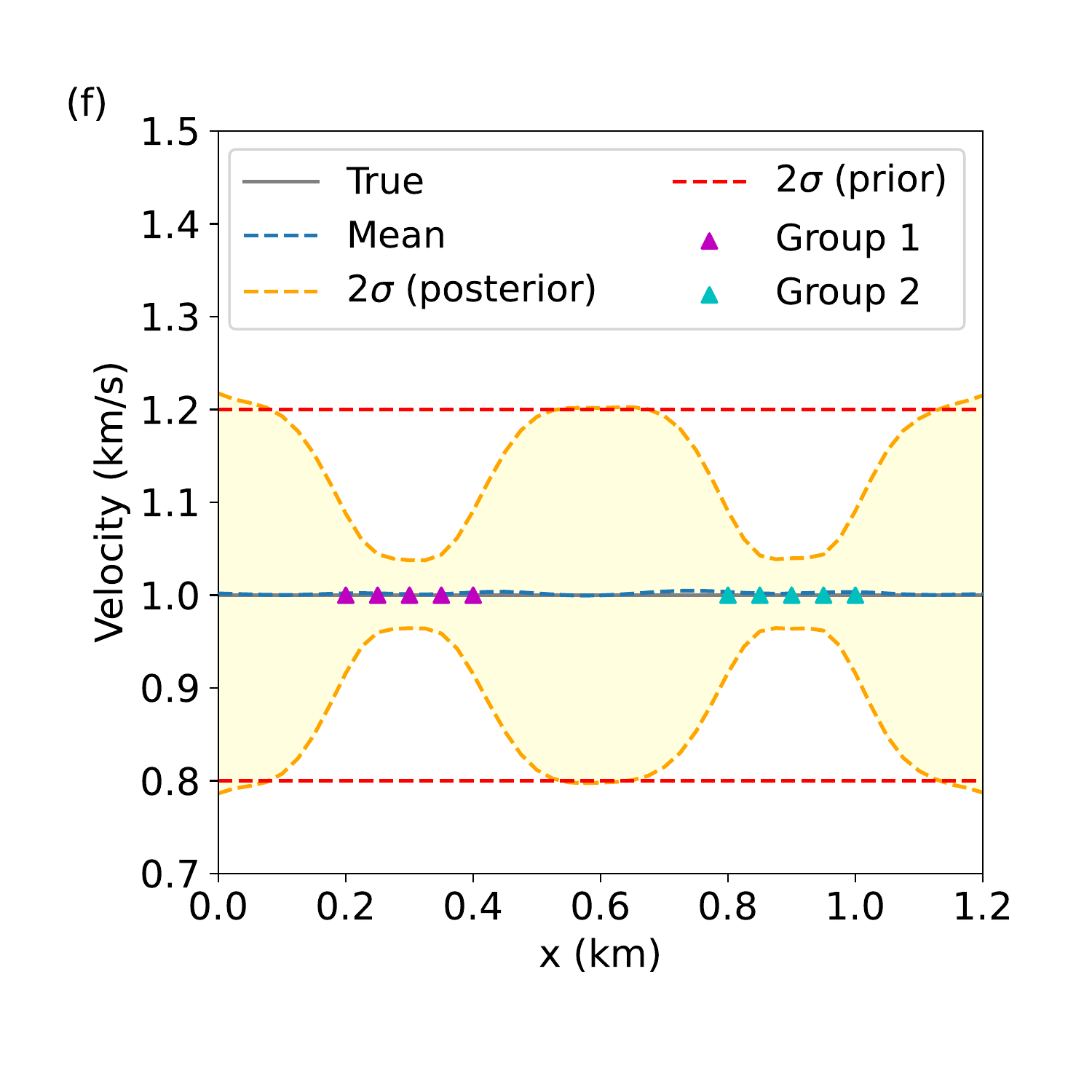}
\end{tabular}
\end{small}
\end{center}
\caption{The results of 1DST. (a)(b) Those obtained by using linearized tomography, which we consider as the ground truth, and vSVGD-PINN-ST, respectively, with $\sigma_2=0.25\,{\rm km}$. 
(c)(d) Those with $\sigma_2=0.15\,{\rm km}$. 
(e)(f) Those with $\sigma_2=0.075\,{\rm km}$. 
}
\label{fig:1D_cl0.25}
\end{figure*}

\begin{figure*}
\begin{center}
\begin{small}
\begin{tabular}{cc}
\includegraphics[clip, width=8cm, bb = 27 42 403 403]{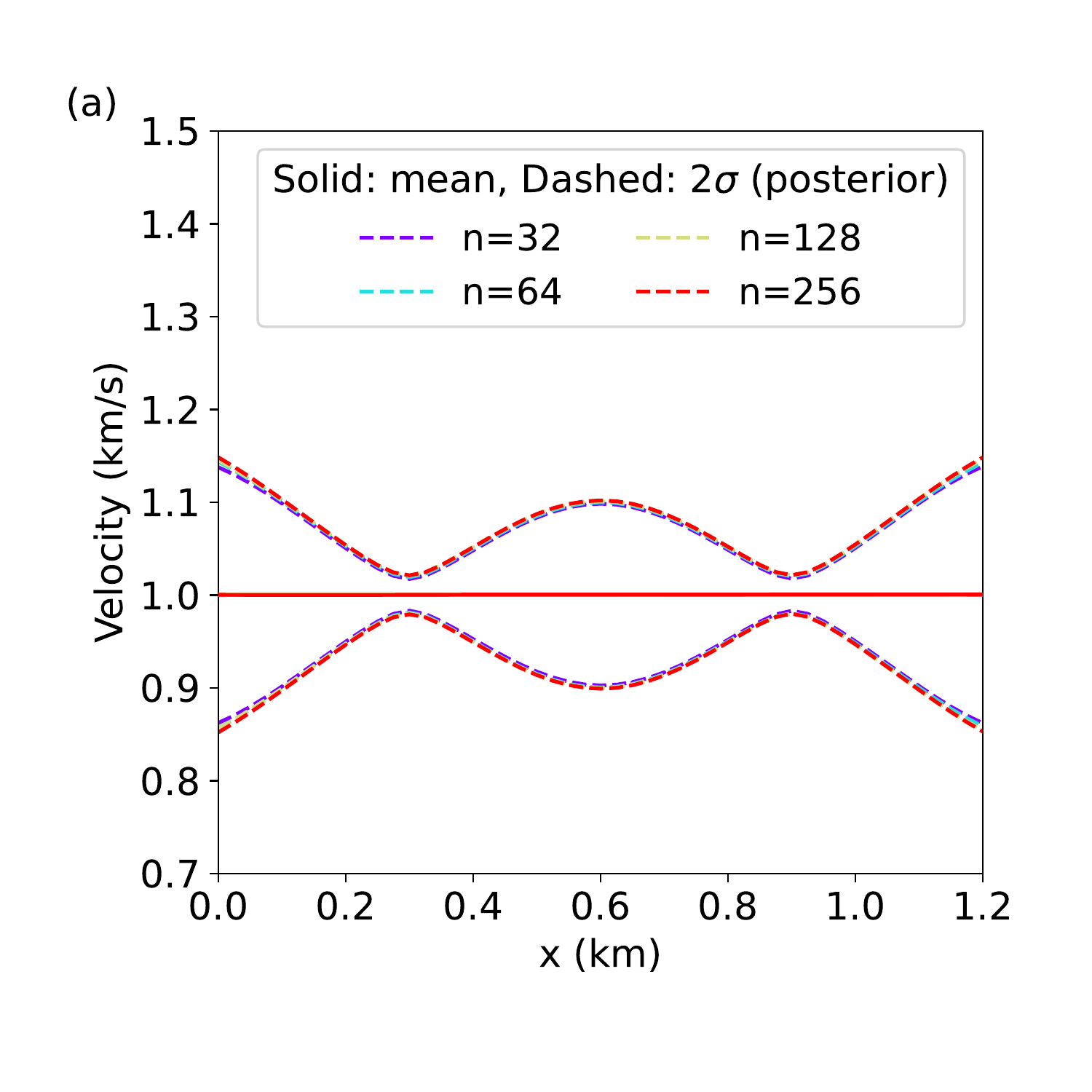}&
\includegraphics[clip, width=8cm, bb = 27 42 403 403]{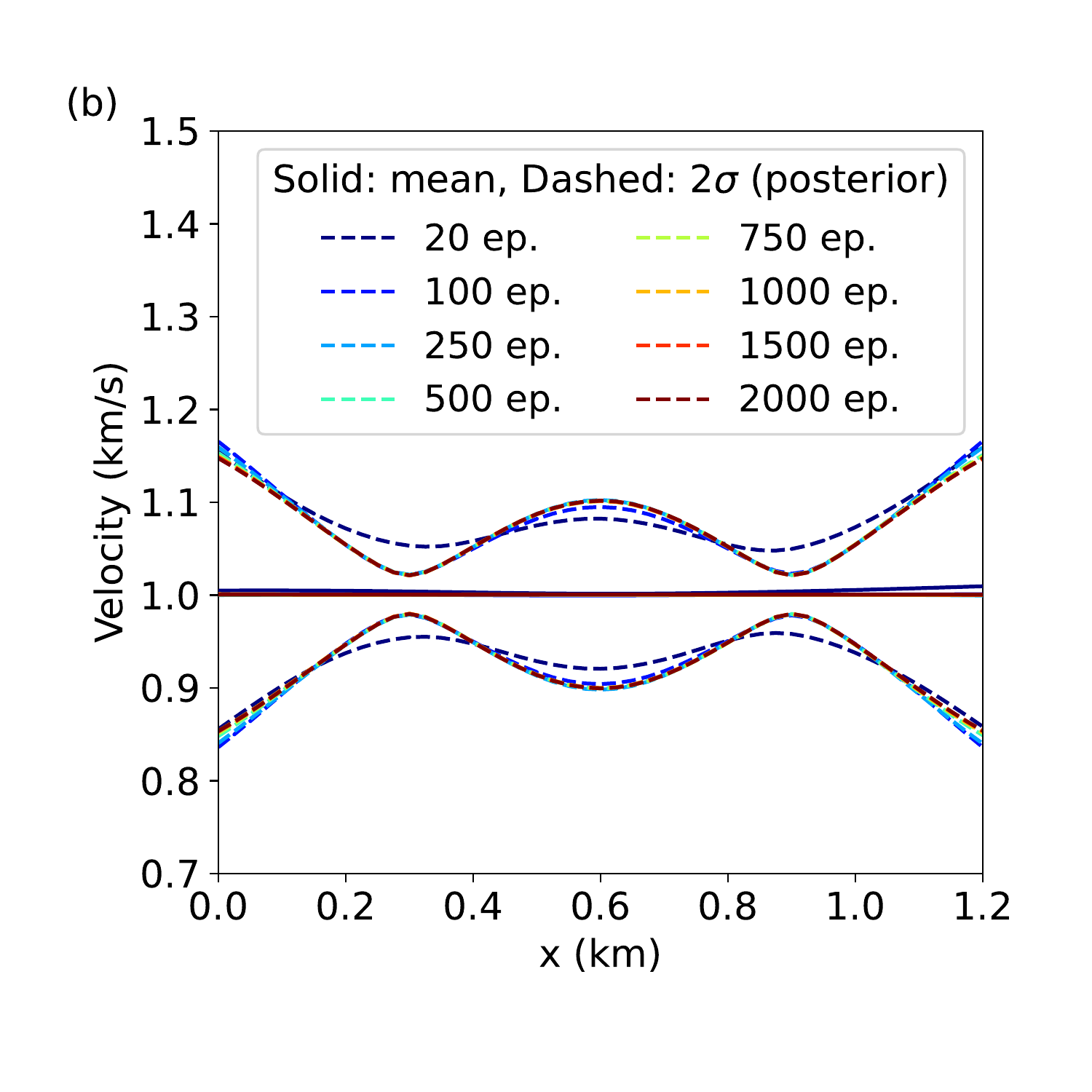}
\end{tabular}
\end{small}
\end{center}
\caption{The relation between the result of 1DST obtained by using vSVGD-PINN-ST with $\sigma_2=0.25\,{\rm km}$ and user-defined parameters. (a) The relation with the number of SVGD particles. (b) That with the number of epochs. }
\label{fig:1D_parameters}
\end{figure*}

%\begin{figure*}
%\begin{center}
%\begin{small}
%\begin{tabular}{cc}
%\includegraphics[clip, width=8cm, bb = 27 42 403 403]{./ST0-SVGD_hp0/velmodel_mean.pdf}&
%\includegraphics[clip, width=8cm, bb = 27 42 403 403]{./ST0-SVGD_hp1/velmodel_mean.pdf}
%\end{tabular}
%\end{small}
%\end{center}
%\caption{The results of 1DST obtained by using SVGD-PINN-ST. Note that the 2-$\sigma$ line of prior probability is not drawn in the scale of velocity because it is given in weight space.  (a) The result obtained with $\sigma_{\theta}^2=10^0$. 
%(b) That with $\sigma_{\theta}^2=10^2$. }
%\label{fig:1D_SVGD}
%\end{figure*}

\begin{figure}
\begin{center}
\begin{small}
\includegraphics[clip, width=8cm, bb = 27 42 403 403]{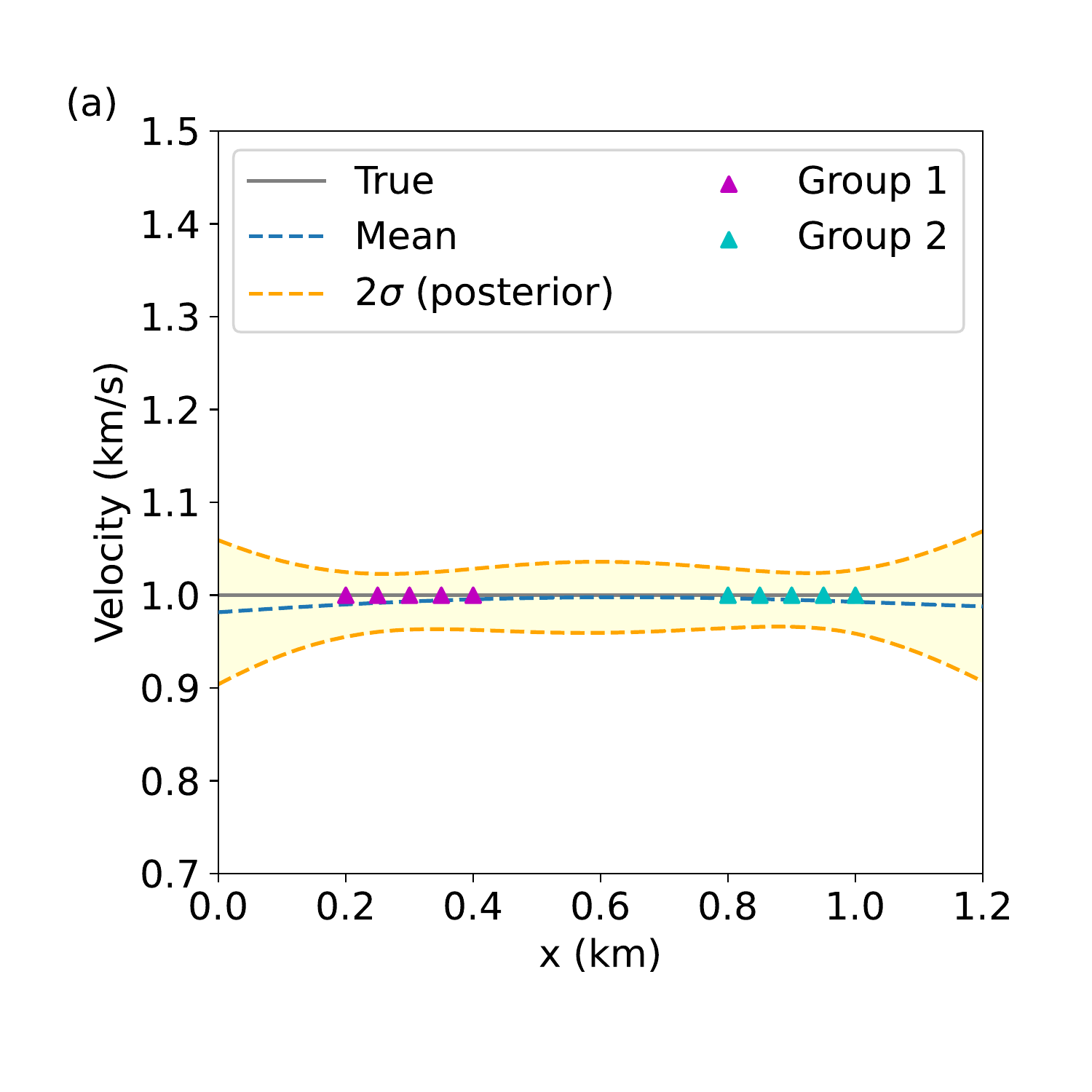}
\end{small}
\end{center}
\caption{The results of 1DST obtained by using SVGD-PINN-ST  with $\sigma_{\theta}^2=10^0$. Note that the 2-$\sigma$ line of prior probability is not drawn in the scale of velocity because it is given in weight space.}
\label{fig:1D_SVGD}
\end{figure}

%\begin{figure}
%\begin{center}
%\begin{small}
%\begin{tabular}{cc}
%\includegraphics[clip, width=8cm, bb = 27 42 403 403]{./ST0-cl0.15-lin_velmodel_mean.pdf}&
%\includegraphics[clip, width=8cm, bb = 27 42 403 403]{./ST0-cl0.15-velmodel_mean.pdf}
%\end{tabular}
%\end{small}
%\end{center}
%\caption{The results of Synthetic Test 0 with $\sigma_2=0.15\,{\rm km}$.  Note that the true value is 1\,km/s for the whole domain, which we do not plot to keep the figure simple. (a) The result obtained by using linearized tomography, which we consider as the ground truth. (b) That obtained by using vSVGD-PINN. }
%\label{fig:1D_cl0.15}
%\end{figure}

\begin{figure*}
\begin{center}
\begin{small}
\begin{tabular}{cc}
\includegraphics[clip, width=6cm, bb = 0 0 350 290]{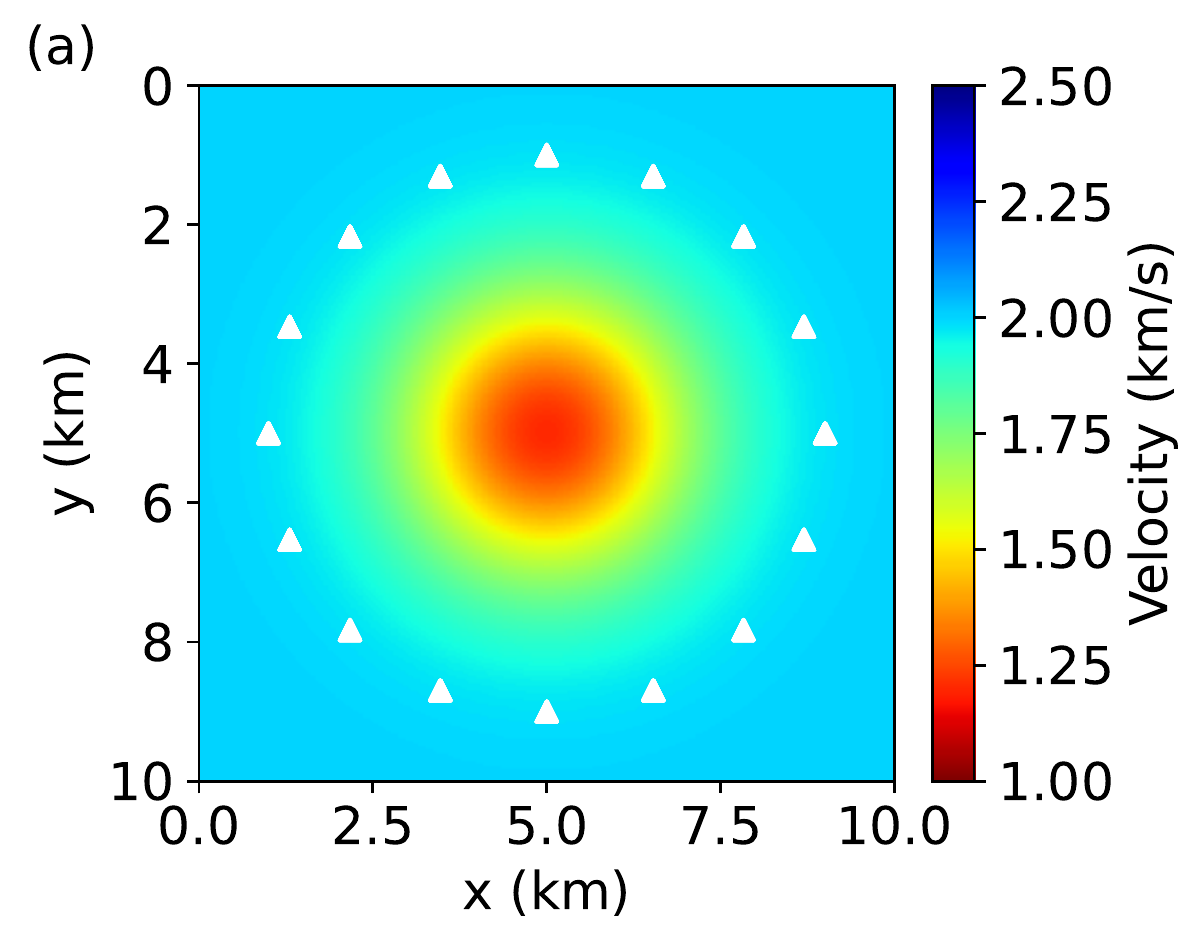}&
\includegraphics[clip, width=6cm, bb = 0 0 350 290]{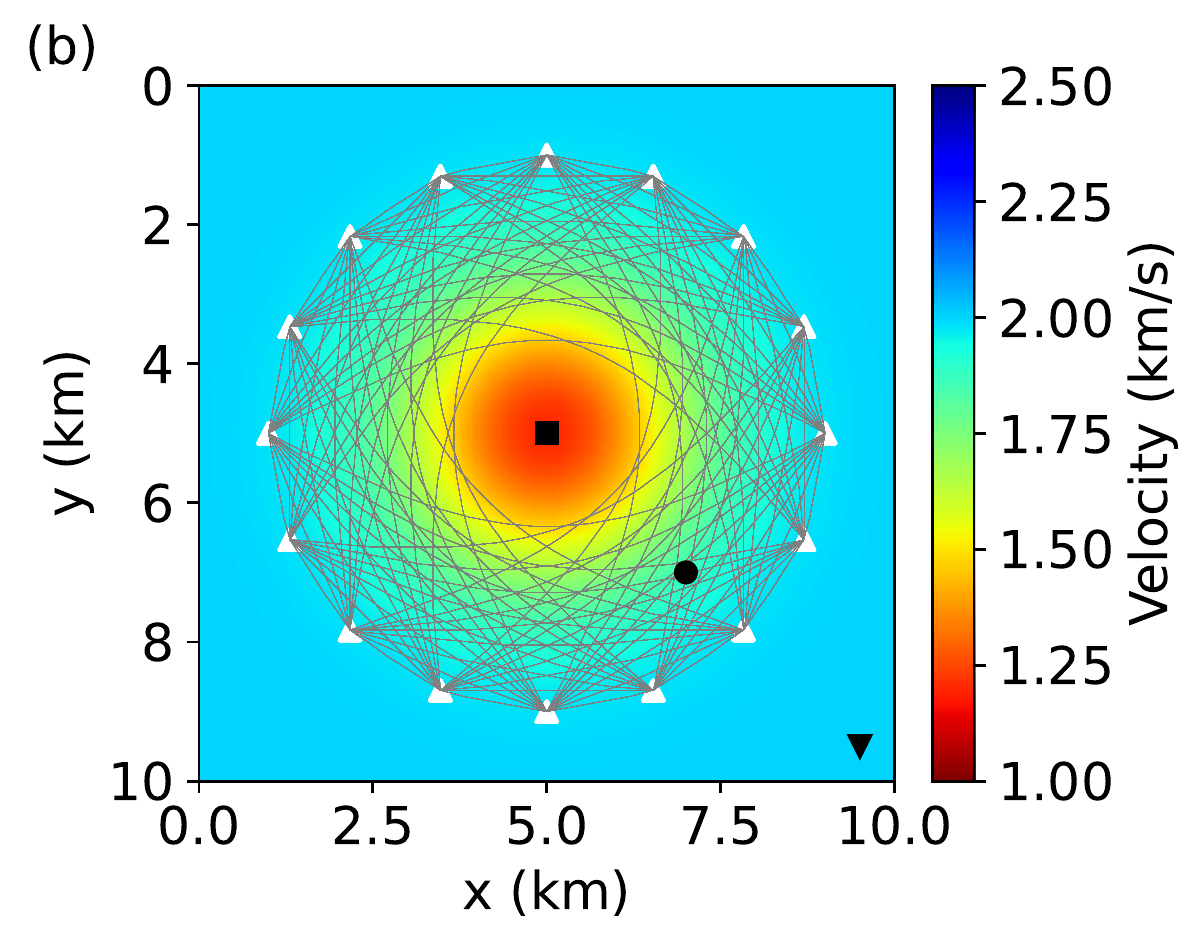}\\
\end{tabular}
\end{small}
\end{center}
\caption{(a) True velocity model in 2DST1. White triangles denote locations of points serving as both sources and receivers. (b) Ray paths (gray lines) between the sources and receiver calculated for the true velocity model. Black marks denotes points to show estimated velocity histograms in Figure \ref{fig:circular_hist}.}
\label{fig:circular}
\end{figure*}

\begin{figure*}
\begin{center}
\begin{small}
\begin{tabular}{ccc}
\includegraphics[clip, width=6cm, bb = 0 0 350 290]{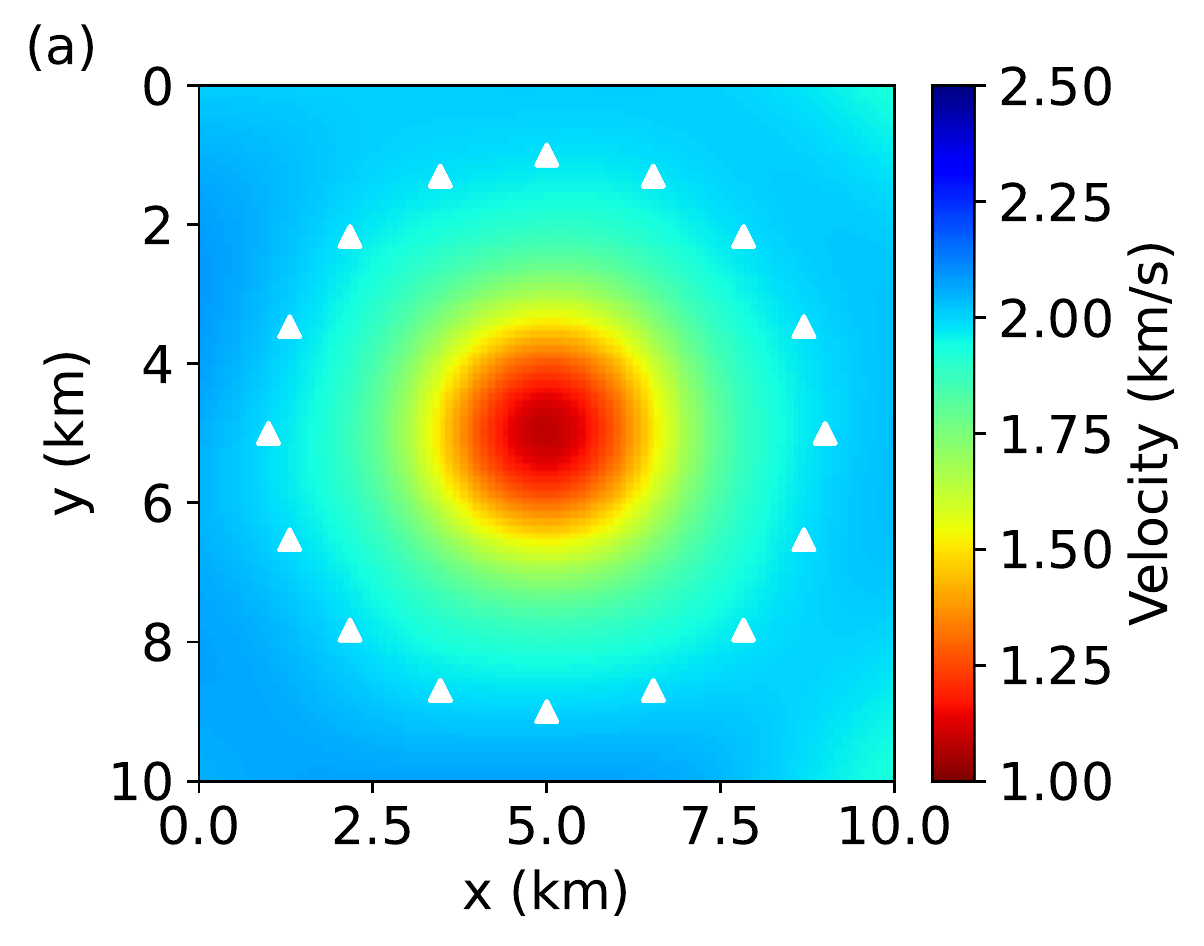}&
\includegraphics[clip, width=6cm, bb = 0 0 350 290]{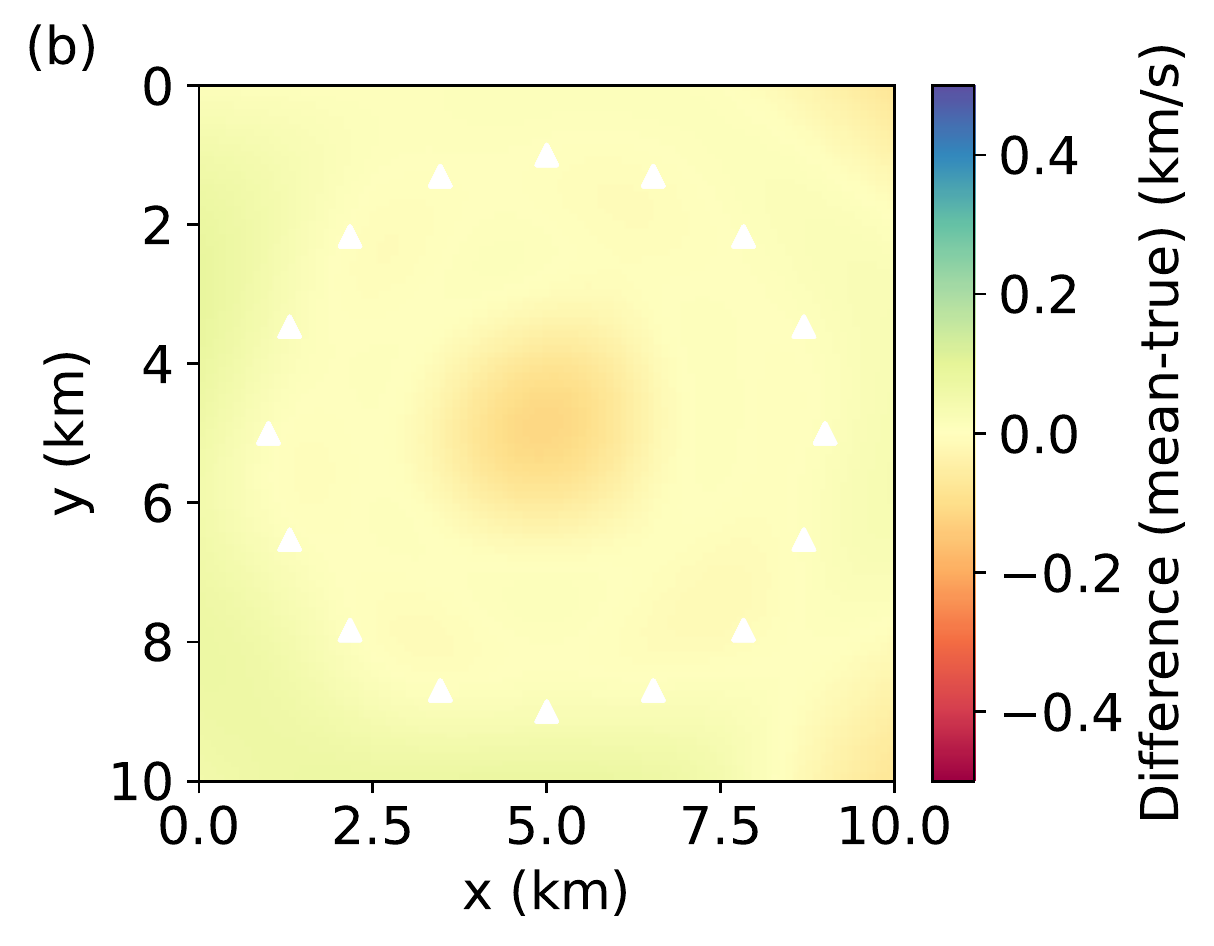}%&
\includegraphics[clip, width=6cm, bb = 0 0 350 290]{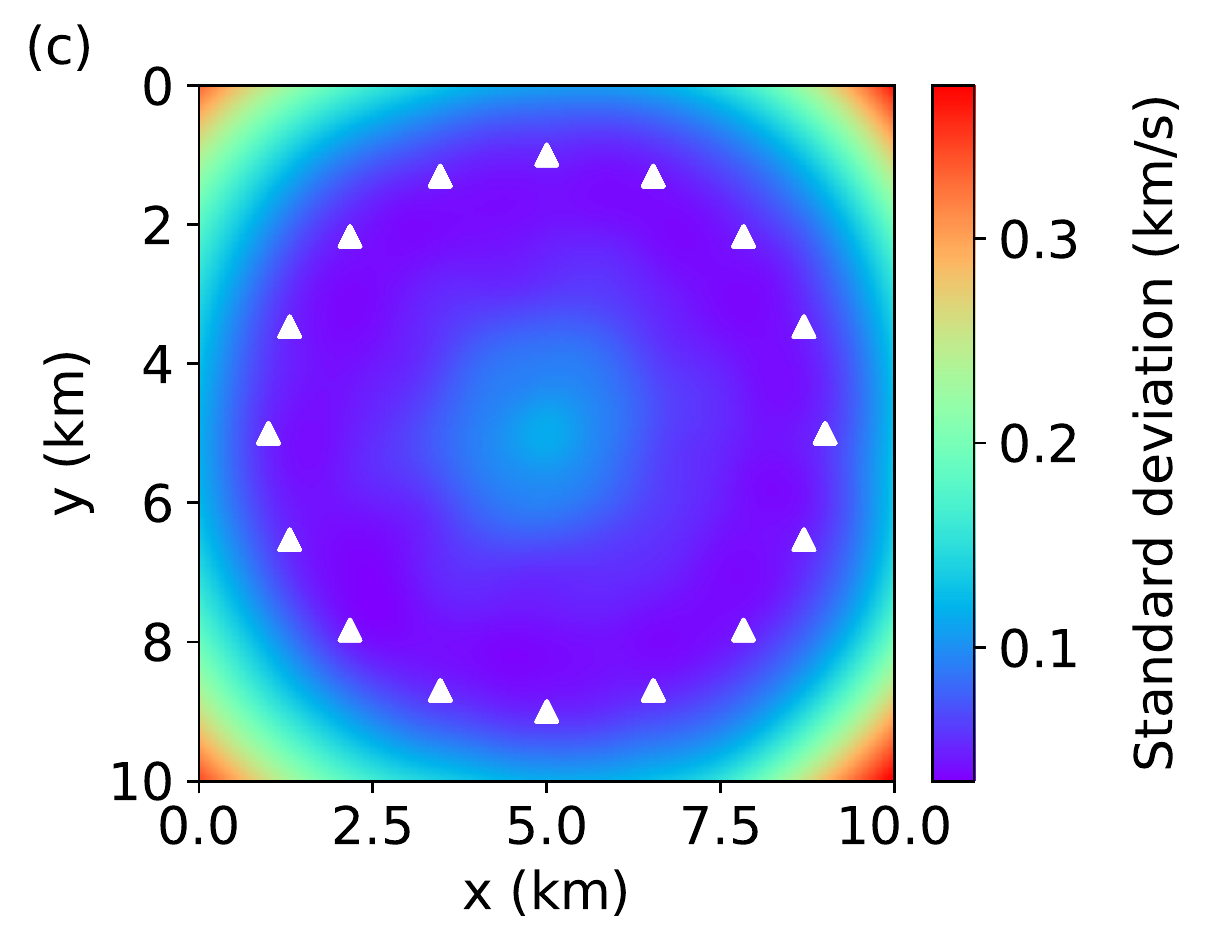}
\end{tabular}
\end{small}
\end{center}
\caption{Estimated velocity models using vSVGD-PINN-ST in 2DST1. White triangles denote locations of points serving as both sources and receivers. 
(a) The mean velocity of the posterior PDF. (b) Difference between the mean and the true model. (c) The standard deviation of the marginal posterior PDF. }
\label{fig:circular_result}
\end{figure*}

%\begin{figure}
%\begin{center}
%%\begin{small}
%\begin{tabular}{ccc}
%\includegraphics[clip, width=6cm, bb = 0 0 350 290]{./ParVI/velmodel_mean.pdf}&
%\includegraphics[clip, width=6cm, bb = 0 0 350 290]{./ParVI/velmodel_diff.pdf}&
%\includegraphics[clip, width=6cm, bb = 0 0 350 290]{./ParVI/velmodel_std.pdf}
%\end{tabular}
%%\end{small}
%\end{center}
%\caption{Estimated velocity models using SVGD. White triangles denote locations of points serving as both sources and receivers. (a) The mean velocity of the posterior PDF. (b) Difference between the mean and the true model. (c) The standard deviation the marginal posterior PDF. }
%\label{fig:circular_result_SVGD}
%\end{figure}

\begin{figure*}
\begin{center}
\begin{small}
\begin{tabular}{ccc}
\includegraphics[clip, width=5.5cm, bb = 0 0 310 310]{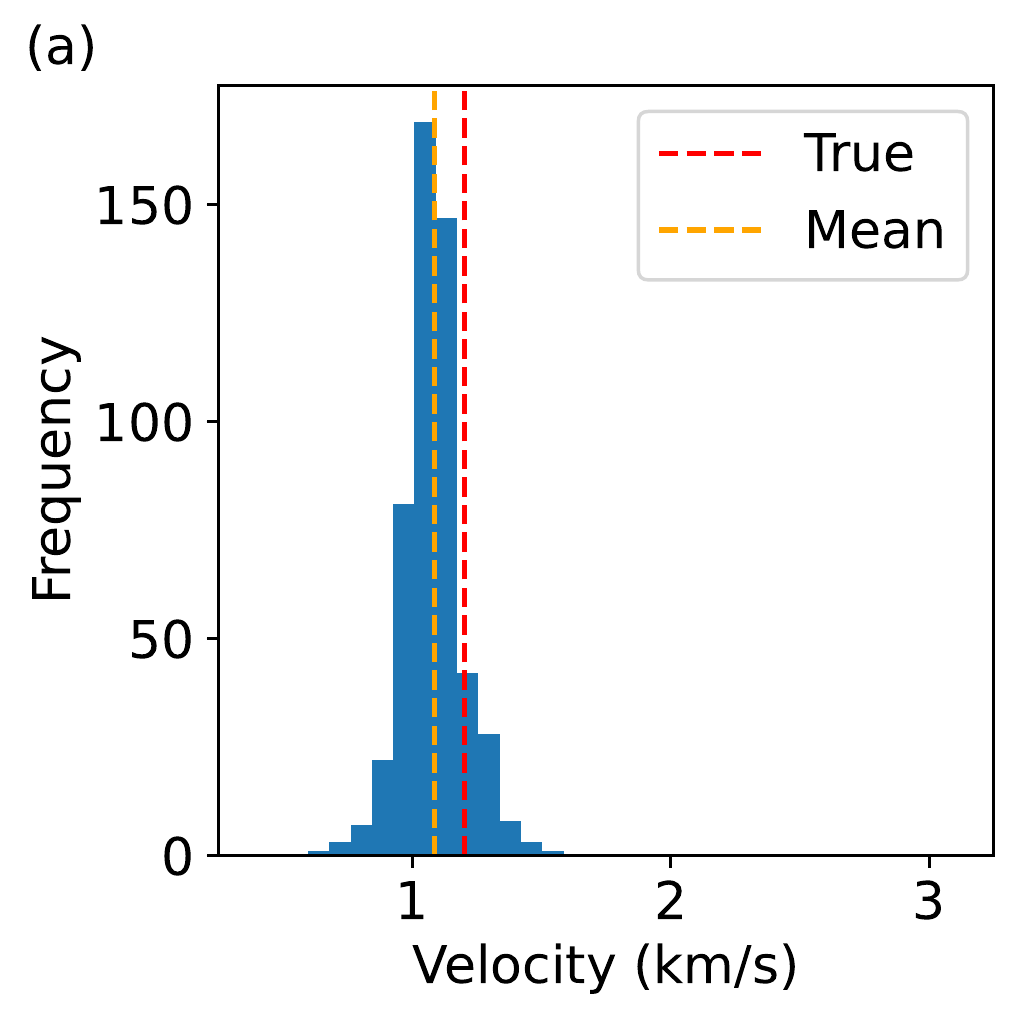}&
\includegraphics[clip, width=5.5cm, bb = 0 0 310 310]{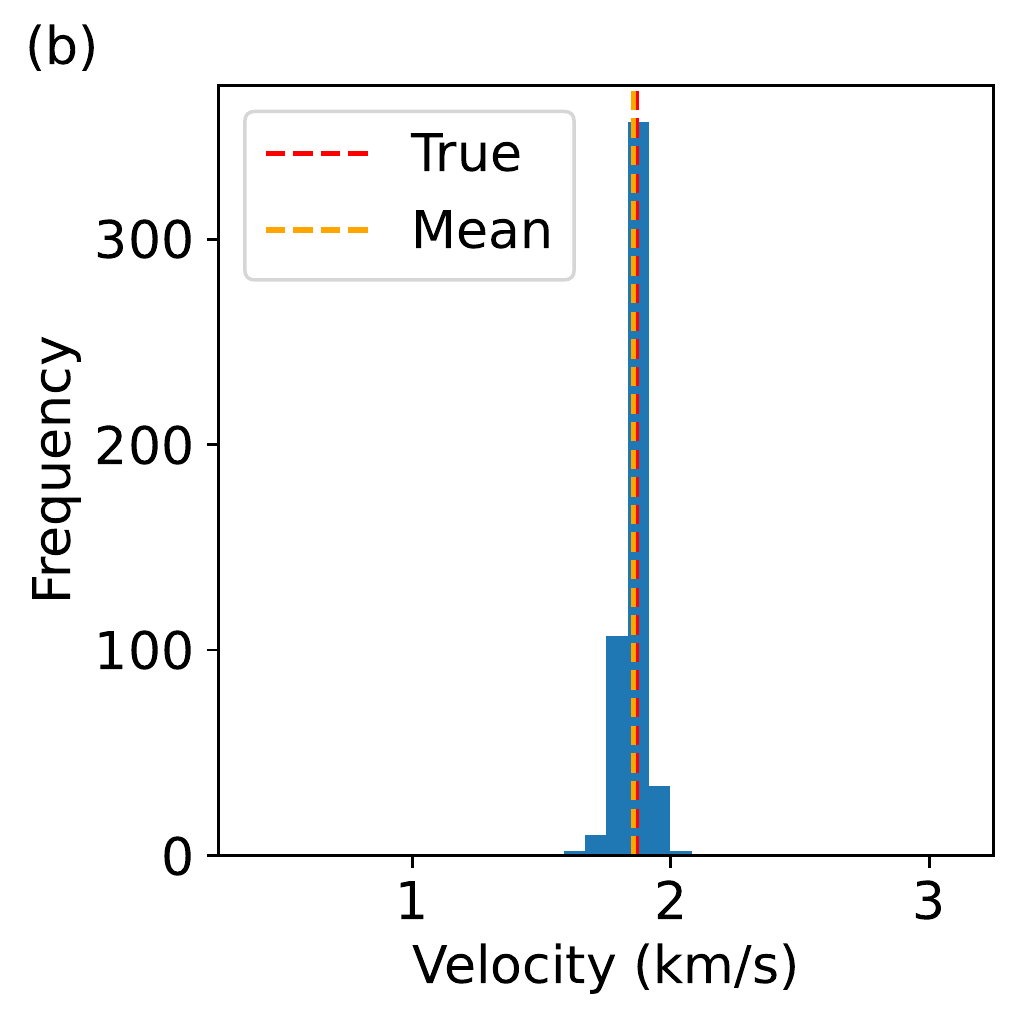}&
\includegraphics[clip, width=5.5cm, bb = 0 0 310 310]{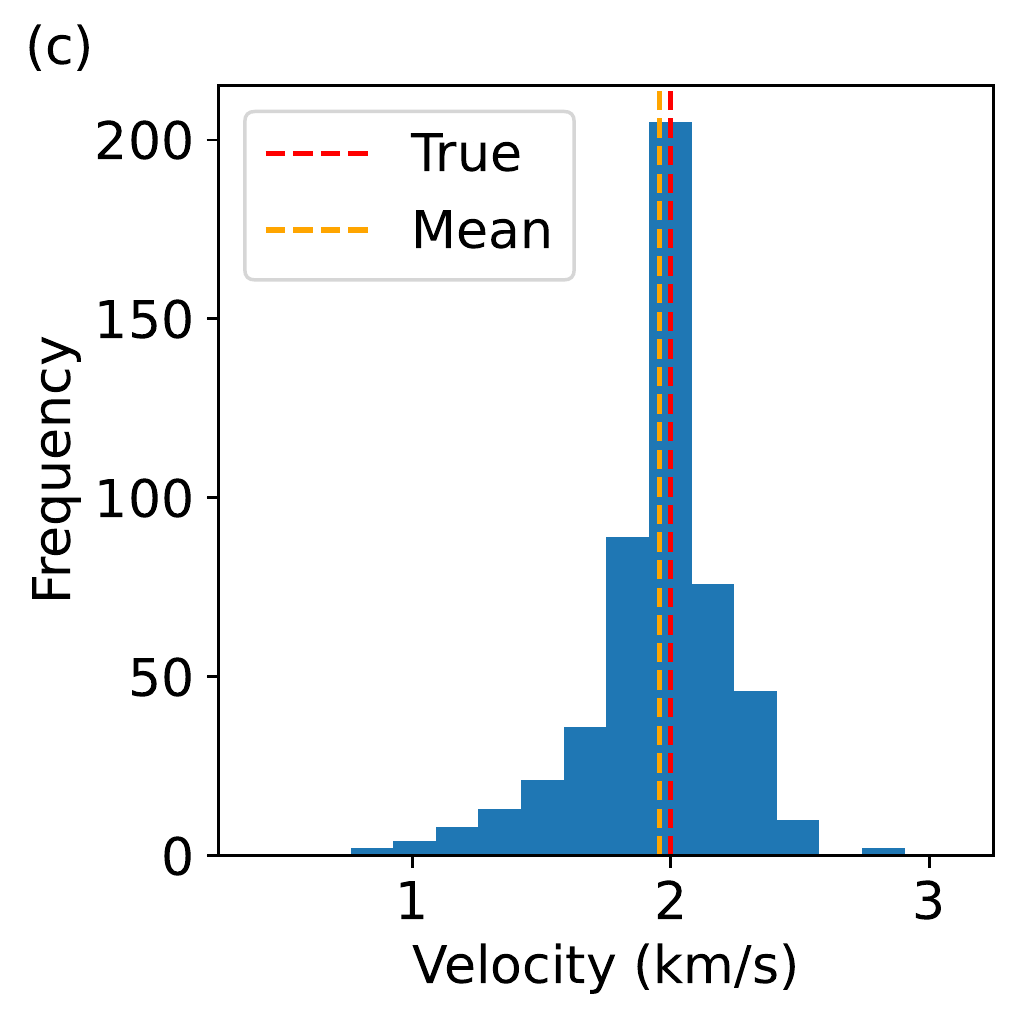}
\end{tabular}
\end{small}
\end{center}
\caption{Histograms of the posterior PDF estimated by using vSVGD-PINN-ST, the estimated mean and the true velocity at the points with black marks in Figure \ref{fig:circular} (b). (a), (b) and (c) are for the square, circle and inverse triangle, respectively. }
\label{fig:circular_hist}
\end{figure*}

%\begin{figure}
%\begin{center}
%\begin{small}
%\begin{tabular}{ccc}
%\includegraphics[clip, width=5.5cm, bb = 0 0 310 310]{./ParVI/hist_000.pdf}&
%\includegraphics[clip, width=5.5cm, bb = 0 0 310 310]{./ParVI/hist_001.pdf}&
%\includegraphics[clip, width=5.5cm, bb = 0 0 310 310]{./ParVI/hist_002.pdf}
%\end{tabular}
%\end{small}
%\end{center}
%\caption{Histograms of the posterior PDF estimated by using SVGD and the true velocity at the points marked in Figure \ref{fig:circular} (b). (a), (b) and (c) are for the square, circle and inverse triangle, respectively. }
%\label{fig:circular_hist_SVGD}
%\end{figure}

\begin{figure*}
\begin{center}
\begin{small}
\includegraphics[clip, width=16cm, bb = 0 0 668 152]{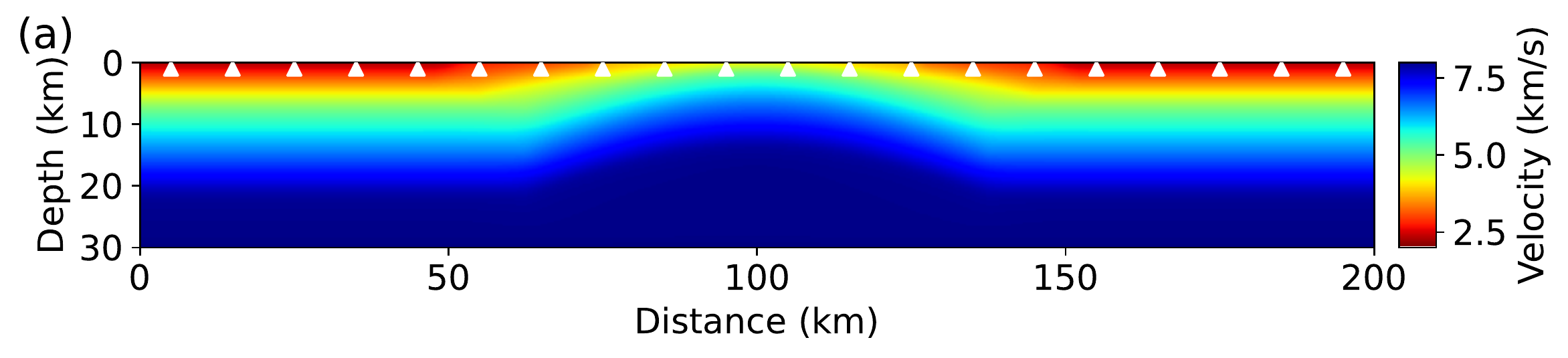}\\
\includegraphics[clip, width=16cm, bb = 0 0 668 152]{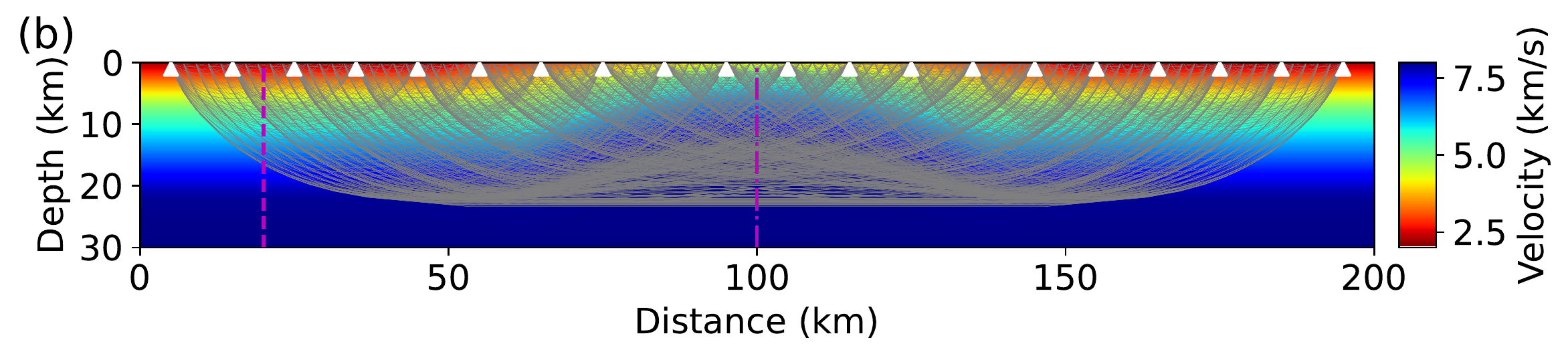}\\
\end{small}
\end{center}
\caption{(a) True velocity model in 2DST2. White triangles denote locations of points serving as sources. (b) Ray paths (gray lines) between the sources and receivers calculated for the true velocity model. Magenta dashed and dotted lines denote the locations to show the line profiles and frequency maps of the estimated velocity in Figure \ref{fig:refraction_hist}. 
%(c) The mean velocity function $\mu({\bf x})$ of the prior Gaussian process. 
}
\label{fig:refraction}
\end{figure*}

\begin{figure*}
\begin{center}
\begin{small}
\includegraphics[clip, width=16cm, bb = 0 0 668 152]{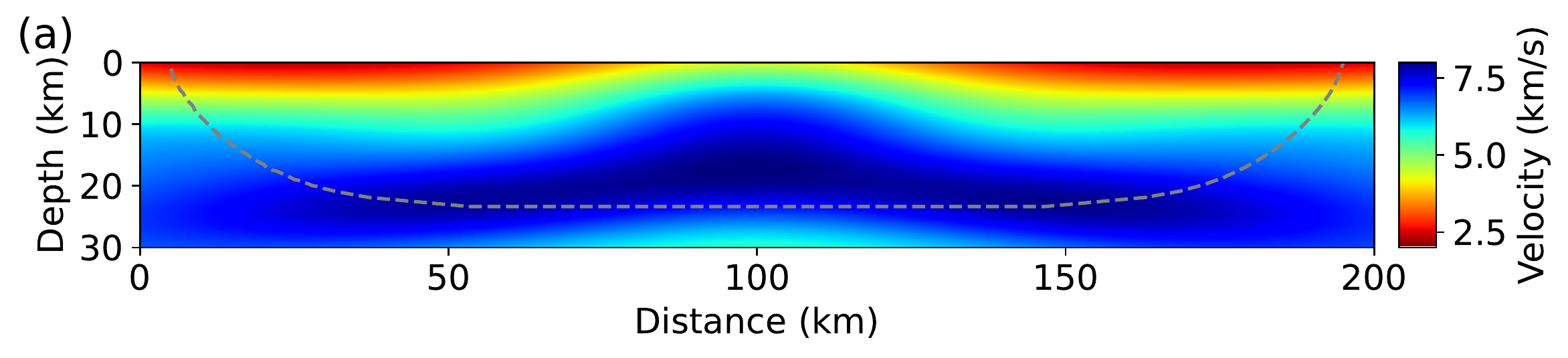}\\
\includegraphics[clip, width=16cm, bb = 0 0 668 152]{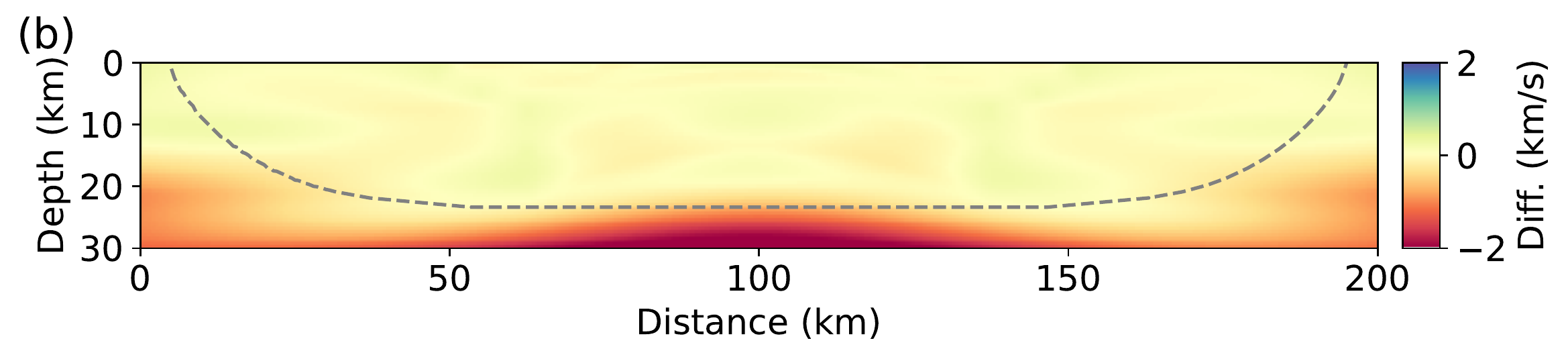}\\
\includegraphics[clip, width=16cm, bb = 0 0 668 152]{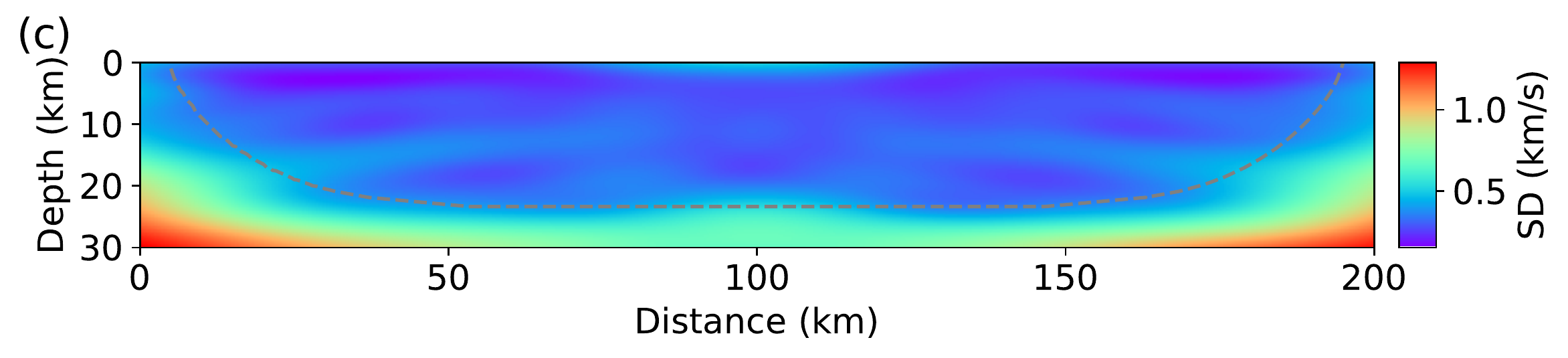}
\end{small}
\end{center}
\caption{Estimated velocity models using vSVGD-PINN-ST in 2DST2. 
White triangles and dashed gray lines denote locations of points serving as sources and the bottom of ray coverage for the true velocity model, respectively. (a) The mean velocity of the posterior PDF. (b) Difference between the mean and the true model. (c) The standard deviation the marginal posterior PDF. }
\label{fig:refraction_result}
\end{figure*}

\begin{figure*}
\begin{center}
\begin{small}
\begin{tabular}{cc}
\includegraphics[clip, width=8cm, bb = 0 0 395 351]{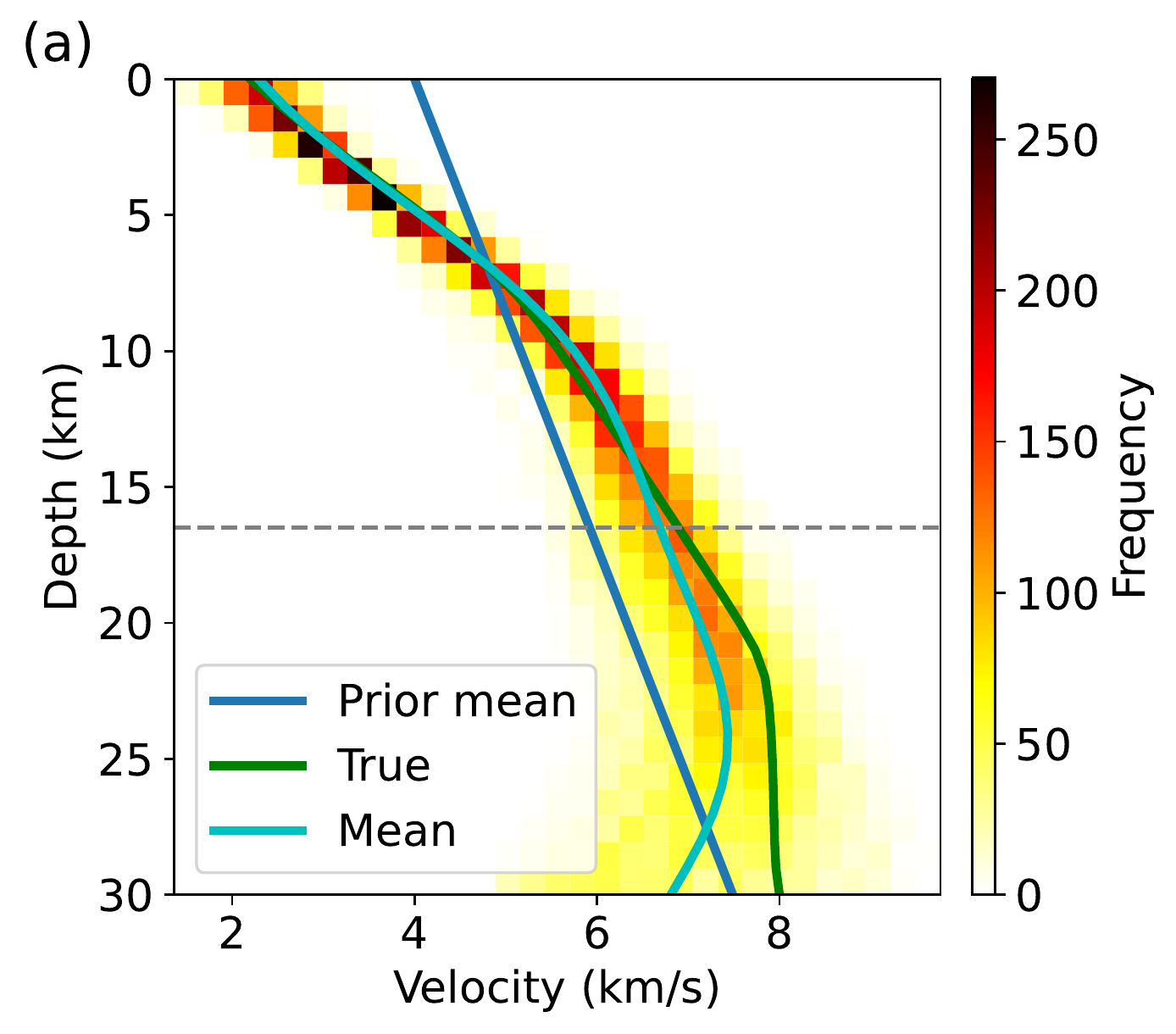}&
\includegraphics[clip, width=8cm, bb = 0 0 395 351]{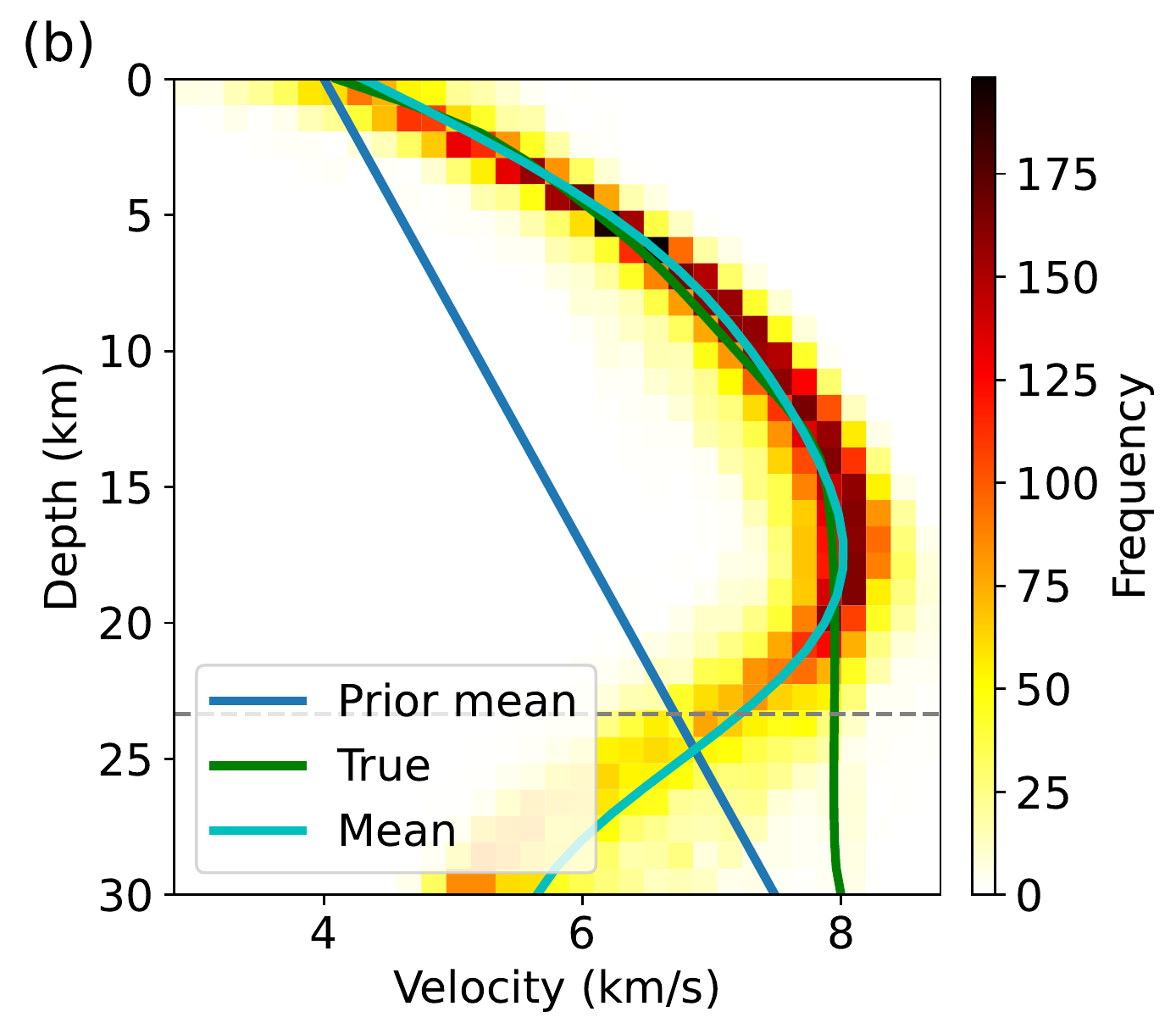}
\end{tabular}
\end{small}
\end{center}
\caption{The line profiles of the mean, true and prior mean velocity and the frequency maps of the posterior PDF estimated using vSVGD-PINN-ST in 2DST2. Dashed gray lines denote the depth of the bottom of ray coverage for the true velocity model. 
(a)(b) Those on the white magenta dashed and dot-dashed lines in Figure \ref{fig:refraction} (b), respectively.}
\label{fig:refraction_hist}
\end{figure*}

%% file: tab.tex
%\begin{table}
%\caption{*: We show only by a simple regression problem (see Supplement material). **: From literature (e.g., \cite{Sun2019}). }
%\label{tab:bnn}
%\begin{center}
%\begin{small}
%\begin{tabular}{cccc} 
%& HMC & ParVI & fParVI \\
%\hline
%Accuracy & $\circledcirc$* & $\bigtriangleup$ & $\bigcirc$ \\
%Time to solution & $\bigtriangleup$* & $\bigcirc$ & $\bigcirc$ \\
%Functional prior & $\bigtriangleup$** & $\bigtriangleup$** & $\circledcirc$ \\
%\end{tabular}
%\end{small}
%\end{center}
%\end{table}

%\begin{table}
%\caption{}
%\label{tab:nn_geom}
%\begin{center}
%\begin{small}
%\begin{tabular}{cccc} 
%& \# hidden  & \# hidden & Activation \\
%& layers & units & function \\
%\hline
%%$f_{{\rm inv}\tau}$ & $\circledcirc$* & $\bigtriangleup$ & $\bigcirc$ \\
%$f_{\rm T}$ & 6 & 100 & Swish \\
%$f_{\rm v}$ & 6 & 20 & Swish \\
%\end{tabular}
%\end{small}
%\end{center}
%\end{table}

%\begin{table}
%\caption{}
%\label{tab:nn_geom}
%\begin{center}
%\begin{small}
%\begin{tabular}{cccc} 
%& \# hidden  & \# hidden & Activation \\
%& layers & units & function \\
%\hline
%%$f_{{\rm inv}\tau}$ & $\circledcirc$* & $\bigtriangleup$ & $\bigcirc$ \\
%$f_{\rm T}$ & 10 & 32 & Swish \\
%$f_{\rm v}$ & 10 & 16 & Swish \\
%\end{tabular}
%\end{small}
%\end{center}
%\end{table}

\begin{table}
\caption{Comparison of the parameters of the kernel function used in the prior Gaussian process in 2DST2 and those estimated based on the maximum likelihood estimation using 1,000 samples of weight parameters of the NN following i.i.d. Gaussian prior. Two cases of the standard deviations the prior were considered, namely, $\sigma_{\theta}=10^{-0.6}$ and $10^{-0.4}$.}
\label{tab:correlation_length}
\begin{center}
\begin{small}
\begin{tabular}{cccc} 
& $\sigma_1^{(*)}$ & $\sigma_x^{(*)}$ & $\sigma_z^{(*)}$ \\
\hline
$\sigma_{\theta}=10^{-0.6}$ & 1.1 & 450.8 & 68.4\\
%$\sigma_{\theta}=10^{-0.5}$ & 1.6 & 127.3 & 14.9\\
$\sigma_{\theta}=10^{-0.4}$ & 2.3 & 23.8 & 4.1\\
Prior in 2DST2 & 1.0 & 35.0 & 10.0\\
\end{tabular}
\end{small}
\end{center}
\end{table}

%% file: app1.tex
\appendix[Normalization strategy of NN input and output]
\label{app:normalization}

In this section, we outline the methods used to normalize the input and output value of the model NNs to improve their convergence performance, using a similar approach to the one in \cite{Rasht-Behesht2022}. 
In order to obtain normalized coordinates in the interval $[-1,1]$ we used the maximum and minimum range values to define the following transformation \cite{Rasht-Behesht2022}: 
\begin{eqnarray}
X\longrightarrow\frac{2X}{{\rm max}(X)-{\rm min}(X)}-1.
\end{eqnarray} 
We apply an additional operator to the output value of $f_{v_{\rm ptb}}$ and $f_{\tau^{-1}}$ defined by
\begin{eqnarray}
a\longrightarrow\frac{\tanh a+1}{2}(a^{\rm max}-a^{\rm min})+a^{\rm min}, 
\end{eqnarray}
where $a$ is the output of either $f_{v_{\rm ptb}}$ or $f_{\tau^{-1}}$ and $ a^{\rm max}$ and $ a^{\rm min}$ represent the maximum and minimum range values defined in the following. 
For $f_{v_{\rm ptb}}$, $a^{\rm max}=v_{\rm ptb}^{\rm max}$ and $a^{\rm min}=v_{\rm ptb}^{\rm min}$, which are given a priori.  
For $f_{\tau^{-1}}$, $a^{\rm max}=v^{\rm max}$ and $a^{\rm min}=v^{\rm min}$, where $v^{\rm max}$ and $v^{\rm min}$ given by 
$v^{\rm max}={\rm max}(v_0({\bf x}))+v_{\rm ptb}^{\rm max}$ and $v^{\rm min}={\rm min}(v_0({\bf x}))+v_{\rm ptb}^{\rm min}$, respectively. 
This operation imposes direct output values of the networks that are included in the interval $[-1,1]$, ensuring that the final upper and lower output limit values are determined by $a^{\rm max}$ and $a^{\rm min}$.

%% file: IEEE_style.bbl

%% file: ms.bbl
\begin{thebibliography}{10}
\providecommand{\url}[1]{#1}
\csname url@samestyle\endcsname
\providecommand{\newblock}{\relax}
\providecommand{\bibinfo}[2]{#2}
\providecommand{\BIBentrySTDinterwordspacing}{\spaceskip=0pt\relax}
\providecommand{\BIBentryALTinterwordstretchfactor}{4}
\providecommand{\BIBentryALTinterwordspacing}{\spaceskip=\fontdimen2\font plus
\BIBentryALTinterwordstretchfactor\fontdimen3\font minus
  \fontdimen4\font\relax}
\providecommand{\BIBforeignlanguage}[2]{{%
\expandafter\ifx\csname l@#1\endcsname\relax
\typeout{** WARNING: IEEEtran.bst: No hyphenation pattern has been}%
\typeout{** loaded for the language `#1'. Using the pattern for}%
\typeout{** the default language instead.}%
\else
\language=\csname l@#1\endcsname
\fi
#2}}
\providecommand{\BIBdecl}{\relax}
\BIBdecl

\bibitem{Bodin2009}
T.~Bodin and M.~Sambridge, ``Seismic tomography with the reversible jump
  algorithm,'' \emph{Geophysical Journal International}, vol. 178, no.~3, pp.
  1411--1436, 2009.

\bibitem{Bodin2012}
T.~Bodin, M.~Sambridge, N.~Rawlinson, and P.~Arroucau, ``Transdimensional
  tomography with unknown data noise,'' \emph{Geophysical Journal
  International}, vol. 189, no.~3, pp. 1536--1556, 2012.

\bibitem{Galetti2015}
E.~Galetti, A.~Curtis, G.~A. Meles, and B.~Baptie, ``{Uncertainty loops in
  travel-time tomography from nonlinear wave physics},'' \emph{Physical review
  letters}, vol. 114, no.~14, p. 148501, 2015.

\bibitem{Zhang2020Seismic}
X.~Zhang and A.~Curtis, ``Seismic tomography using variational inference
  methods,'' \emph{Journal of Geophysical Research: Solid Earth}, vol. 125,
  no.~4, p. e2019JB018589, 2020.

\bibitem{Ryberg2018}
T.~Ryberg and C.~Haberland, ``Bayesian inversion of refraction seismic
  traveltime data,'' \emph{Geophysical Journal International}, vol. 212, no.~3,
  pp. 1645--1656, 2018.

\bibitem{Piana2015}
N.~Piana~Agostinetti, G.~Giacomuzzi, and A.~Malinverno, ``{Local
  three-dimensional earthquake tomography by trans-dimensional Monte Carlo
  sampling},'' \emph{Geophysical Journal International}, vol. 201, no.~3, pp.
  1598--1617, 2015.

\bibitem{Hawkins2015}
R.~Hawkins and M.~Sambridge, ``Geophysical imaging using trans-dimensional
  trees,'' \emph{Geophysical Journal International}, vol. 203, no.~2, pp.
  972--1000, 2015.

\bibitem{Burdick2017}
S.~Burdick and V.~Leki{\'c}, ``{Velocity variations and uncertainty from
  transdimensional P-wave tomography of North America},'' \emph{Geophysical
  Journal International}, vol. 209, no.~2, pp. 1337--1351, 2017.

\bibitem{Raissi2019}
M.~Raissi, P.~Perdikaris, and G.~E. Karniadakis, ``{Physics-informed neural
  networks: A deep learning framework for solving forward and inverse problems
  involving nonlinear partial differential equations},'' \emph{Journal of
  Computational physics}, vol. 378, pp. 686--707, 2019.

\bibitem{Smith2020}
J.~D. Smith, K.~Azizzadenesheli, and Z.~E. Ross, ``Eikonet: Solving the eikonal
  equation with deep neural networks,'' \emph{IEEE Transactions on Geoscience
  and Remote Sensing}, vol.~59, no.~12, pp. 10\,685--10\,696, 2020.

\bibitem{Waheed2021PINNeik}
U.~B. Waheed, E.~Haghighat, T.~Alkhalifah, C.~Song, and Q.~Hao, ``{PINNeik:
  Eikonal solution using physics-informed neural networks},'' \emph{Computers
  \& Geosciences}, vol. 155, p. 104833, 2021.

\bibitem{Song2022}
C.~Song and Y.~Wang, ``{Simulating seismic multifrequency wavefields with the
  Fourier feature physics-informed neural network},'' \emph{Geophysical Journal
  International}, vol. 232, no.~3, pp. 1503--1514, 2022.

\bibitem{Okazaki2022}
T.~Okazaki, T.~Ito, K.~Hirahara, and N.~Ueda, ``{Physics-informed deep learning
  approach for modeling crustal deformation},'' \emph{Nature Communications},
  vol.~13, no.~1, 2022.

\bibitem{Waheed2021PINNtomo}
U.~B. Waheed, T.~Alkhalifah, E.~Haghighat, C.~Song, and J.~Virieux,
  ``{PINNtomo: Seismic tomography using physics-informed neural networks},''
  \emph{arXiv preprint arXiv:2104.01588}, 2021.

\bibitem{Chen2022}
Y.~Chen, S.~A. de~Ridder, S.~Rost, Z.~Guo, X.~Wu, and Y.~Chen, ``{Eikonal
  Tomography With Physics-Informed Neural Networks: Rayleigh Wave Phase
  Velocity in the Northeastern Margin of the Tibetan Plateau},''
  \emph{Geophysical Research Letters}, vol.~49, no.~21, p. e2022GL099053, 2022.

\bibitem{Rasht-Behesht2022}
M.~Rasht-Behesht, C.~Huber, K.~Shukla, and G.~E. Karniadakis,
  ``{Physics-Informed Neural Networks (PINNs) for Wave Propagation and Full
  Waveform Inversions},'' \emph{Journal of Geophysical Research: Solid Earth},
  vol. 127, no.~5, p. e2021JB023120, 2022.

\bibitem{Gawlikowski2021}
J.~Gawlikowski, C.~R.~N. Tassi, M.~Ali, J.~Lee, M.~Humt, J.~Feng, A.~Kruspe,
  R.~Triebel, P.~Jung, R.~Roscher \emph{et~al.}, ``A survey of uncertainty in
  deep neural networks,'' \emph{arXiv preprint arXiv:2107.03342}, 2021.

\bibitem{Duane1987}
S.~Duane, A.~D. Kennedy, B.~J. Pendleton, and D.~Roweth, ``Hybrid monte
  carlo,'' \emph{Physics letters B}, vol. 195, no.~2, pp. 216--222, 1987.

\bibitem{Yang2021}
L.~Yang, X.~Meng, and G.~E. Karniadakis, ``{B-PINNs: Bayesian physics-informed
  neural networks for forward and inverse PDE problems with noisy data},''
  \emph{Journal of Computational Physics}, vol. 425, p. 109913, 2021.

\bibitem{Linka2022}
K.~Linka, A.~Schafer, X.~Meng, Z.~Zou, G.~E. Karniadakis, and E.~Kuhl,
  ``{Bayesian Physics-Informed Neural Networks for real-world nonlinear
  dynamical systems},'' \emph{arXiv preprint arXiv:2205.08304}, 2022.

\bibitem{Psaros2023}
A.~F. Psaros, X.~Meng, Z.~Zou, L.~Guo, and G.~E. Karniadakis, ``{Uncertainty
  quantification in scientific machine learning: Methods, metrics, and
  comparisons},'' \emph{Journal of Computational Physics}, p. 111902, 2023.

\bibitem{Liu2016}
Q.~Liu and D.~Wang, ``{Stein variational gradient descent: A general purpose
  bayesian inference algorithm},'' \emph{Advances in neural information
  processing systems}, vol.~29, 2016.

\bibitem{Sun2020}
L.~Sun and J.-X. Wang, ``{Physics-constrained Bayesian neural network for fluid
  flow reconstruction with sparse and noisy data},'' \emph{Theoretical and
  Applied Mechanics Letters}, vol.~10, no.~3, pp. 161--169, 2020.

\bibitem{Wang2019}
Z.~Wang, T.~Ren, J.~Zhu, and B.~Zhang, ``{Function Space Particle Optimization
  for Bayesian Neural Networks},'' in \emph{International Conference on
  Learning Representations}, 2019.

\bibitem{Sun2019}
S.~Sun, G.~Zhang, J.~Shi, and R.~Grosse, ``{Functional variational Bayesian
  neural networks},'' in \emph{International Conference on Learning
  Representations}, 2019.

\bibitem{Lewis1985}
\BIBentryALTinterwordspacing
J.~M. Lewis and J.~C. Derber, ``{The use of adjoint equations to solve a
  variational adjustment problem with advective constraints},'' \emph{Tellus
  A}, vol.~37, no.~4, 1985. [Online]. Available:
  \url{http://www.tellusa.net/index.php/tellusa/article/view/11675}
\BIBentrySTDinterwordspacing

\bibitem{Grubas2023}
S.~Grubas, A.~Duchkov, and G.~Loginov, ``{Neural Eikonal solver: Improving
  accuracy of physics-informed neural networks for solving eikonal equation in
  case of caustics},'' \emph{Journal of Computational Physics}, vol. 474, p.
  111789, 2023.

\bibitem{Wang2022}
S.~Wang, X.~Yu, and P.~Perdikaris, ``{When and why PINNs fail to train: A
  neural tangent kernel perspective},'' \emph{Journal of Computational
  Physics}, vol. 449, p. 110768, 2022.

\bibitem{Zhang2020Variational}
X.~Zhang and A.~Curtis, ``Variational full-waveform inversion,''
  \emph{Geophysical Journal International}, vol. 222, no.~1, pp. 406--411,
  2020.

\bibitem{Smith2022}
J.~D. Smith, Z.~E. Ross, K.~Azizzadenesheli, and J.~B. Muir, ``{HypoSVI:
  Hypocentre inversion with Stein variational inference and physics informed
  neural networks},'' \emph{Geophysical Journal International}, vol. 228,
  no.~1, pp. 698--710, 2022.

\bibitem{Kingma2015}
D.~P. Kingma and J.~Ba, ``{Adam: A method for stochastic optimization},'' in
  \emph{International Conference on Learning Representations}, 2015.

\bibitem{Liu1989}
D.~C. Liu and J.~Nocedal, ``{On the limited memory BFGS method for large scale
  optimization},'' \emph{Mathematical programming}, vol.~45, no.~1, pp.
  503--528, 1989.

\bibitem{Warton2008}
D.~I. Warton, ``Penalized normal likelihood and ridge regularization of
  correlation and covariance matrices,'' \emph{Journal of the American
  Statistical Association}, vol. 103, no. 481, pp. 340--349, 2008.

\bibitem{Ramachandran2018}
P.~Ramachandran, B.~Zoph, and Q.~V. Le, ``Searching for activation functions,''
  in \emph{International Conference on Learning Representations}, 2018.

\bibitem{He2015}
K.~He, X.~Zhang, S.~Ren, and J.~Sun, ``{Delving deep into rectifiers:
  Surpassing human-level performance on imagenet classification},'' in
  \emph{Proceedings of the IEEE international conference on computer vision},
  2015, pp. 1026--1034.

\bibitem{Rahaman2019}
N.~Rahaman, A.~Baratin, D.~Arpit, F.~Draxler, M.~Lin, F.~Hamprecht, Y.~Bengio,
  and A.~Courville, ``On the spectral bias of neural networks,'' in
  \emph{International Conference on Machine Learning}.\hskip 1em plus 0.5em
  minus 0.4em\relax PMLR, 2019, pp. 5301--5310.

\bibitem{Zhao2005}
H.~Zhao, ``A fast sweeping method for eikonal equations,'' \emph{Mathematics of
  computation}, vol.~74, no. 250, pp. 603--627, 2005.

\bibitem{Giroux2021}
B.~Giroux, ``ttcrpy: A python package for traveltime computation and
  raytracing,'' \emph{SoftwareX}, vol.~16, p. 100834, 2021.

\bibitem{Liu2019}
L.~Liu, H.~Jiang, P.~He, W.~Chen, X.~Liu, J.~Gao, and J.~Han, ``{On the
  Variance of the Adaptive Learning Rate and Beyond},'' in \emph{Proceedings of
  the Eighth International Conference on Learning Representations (ICLR 2020)},
  April 2020.

\bibitem{Zelt2003}
C.~A. Zelt, K.~Sain, J.~V. Naumenko, and D.~S. Sawyer, ``Assessment of crustal
  velocity models using seismic refraction and reflection tomography,''
  \emph{Geophysical Journal International}, vol. 153, no.~3, pp. 609--626,
  2003.

\bibitem{Korenaga2000}
J.~Korenaga, W.~Holbrook, G.~Kent, P.~Kelemen, R.~Detrick, H.-C. Larsen,
  J.~Hopper, and T.~Dahl-Jensen, ``{Crustal structure of the southeast
  Greenland margin from joint refraction and reflection seismic tomography},''
  \emph{Journal of Geophysical Research: Solid Earth}, vol. 105, no.~B9, pp.
  21\,591--21\,614, 2000.

\bibitem{Kodaira2014}
S.~Kodaira, G.~Fujie, M.~Yamashita, T.~Sato, T.~Takahashi, and N.~Takahashi,
  ``{Seismological evidence of mantle flow driving plate motions at a
  palaeo-spreading centre},'' \emph{Nature Geoscience}, vol.~7, no.~5, pp.
  371--375, 2014.

\bibitem{Bishop2006}
C.~M. Bishop, \emph{Pattern recognition and machine learning}.\hskip 1em plus
  0.5em minus 0.4em\relax Springer, 2006.

\bibitem{Karniadakis2021}
G.~E. Karniadakis, I.~G. Kevrekidis, L.~Lu, P.~Perdikaris, S.~Wang, and
  L.~Yang, ``Physics-informed machine learning,'' \emph{Nature Reviews
  Physics}, vol.~3, no.~6, pp. 422--440, 2021.

\bibitem{Wang2021}
S.~Wang, H.~Wang, and P.~Perdikaris, ``{On the eigenvector bias of fourier
  feature networks: From regression to solving multi-scale pdes with
  physics-informed neural networks},'' \emph{Computer Methods in Applied
  Mechanics and Engineering}, vol. 384, p. 113938, 2021.

\bibitem{Jagtap2020Adaptive}
A.~D. Jagtap, K.~Kawaguchi, and G.~E. Karniadakis, ``Adaptive activation
  functions accelerate convergence in deep and physics-informed neural
  networks,'' \emph{Journal of Computational Physics}, vol. 404, p. 109136,
  2020.

\bibitem{Tancik2020}
M.~Tancik, P.~Srinivasan, B.~Mildenhall, S.~Fridovich-Keil, N.~Raghavan,
  U.~Singhal, R.~Ramamoorthi, J.~Barron, and R.~Ng, ``{Fourier features let
  networks learn high frequency functions in low dimensional domains},''
  \emph{Advances in Neural Information Processing Systems}, vol.~33, pp.
  7537--7547, 2020.

\bibitem{Jagtap2021Extended}
A.~D. Jagtap and G.~E. Karniadakis, ``{Extended Physics-informed Neural
  Networks (XPINNs): A Generalized Space-Time Domain Decomposition based Deep
  Learning Framework for Nonlinear Partial Differential Equations},'' in
  \emph{AAAI Spring Symposium: MLPS}, 2021, pp. 2002--2041.

\bibitem{Wang2022Causality}
S.~Wang, S.~Sankaran, and P.~Perdikaris, ``Respecting causality is all you need
  for training physics-informed neural networks,'' \emph{arXiv preprint
  arXiv:2203.07404}, 2022.

\bibitem{Mao2020}
Z.~Mao, A.~D. Jagtap, and G.~E. Karniadakis, ``Physics-informed neural networks
  for high-speed flows,'' \emph{Computer Methods in Applied Mechanics and
  Engineering}, vol. 360, p. 112789, 2020.

\bibitem{Metropolis1953}
N.~Metropolis, A.~W. Rosenbluth, M.~N. Rosenbluth, A.~H. Teller, and E.~Teller,
  ``Equation of state calculations by fast computing machines,'' \emph{The
  journal of chemical physics}, vol.~21, no.~6, pp. 1087--1092, 1953.

\bibitem{Hastings1970}
W.~K. Hastings, ``{Monte Carlo sampling methods using Markov chains and their
  applications},'' 1970.

\bibitem{Green1995}
P.~J. Green, ``{Reversible jump Markov chain Monte Carlo computation and
  Bayesian model determination},'' \emph{Biometrika}, vol.~82, no.~4, pp.
  711--732, 1995.

\bibitem{Wang2019matrix-valued}
D.~Wang, Z.~Tang, C.~Bajaj, and Q.~Liu, ``{Stein variational gradient descent
  with matrix-valued kernels},'' \emph{Advances in neural information
  processing systems}, vol.~32, 2019.

\bibitem{Ai2022}
Q.~Ai, S.~Liu, L.~He, and Z.~Xu, ``{Stein Variational Gradient Descent with
  Multiple Kernels},'' \emph{Cognitive Computation}, pp. 1--11, 2022.

\bibitem{Detommaso2018}
G.~Detommaso, T.~Cui, Y.~Marzouk, A.~Spantini, and R.~Scheichl, ``{A Stein
  variational Newton method},'' \emph{Advances in Neural Information Processing
  Systems}, vol.~31, 2018.

\bibitem{Zhu2020}
M.~Zhu, C.~Liu, and J.~Zhu, ``{Variance reduction and quasi-Newton for
  particle-based variational inference},'' in \emph{International Conference on
  Machine Learning}.\hskip 1em plus 0.5em minus 0.4em\relax PMLR, 2020, pp.
  11\,576--11\,587.

\bibitem{Tran2022}
B.-H. Tran, S.~Rossi, D.~Milios, and M.~Filippone, ``{All You Need is a Good
  Functional Prior for Bayesian Deep Learning},'' \emph{Journal of Machine
  Learning Research}, vol.~23, no.~74, pp. 1--56, 2022.

\end{thebibliography}
